\documentclass[12pt]{article}
\pdfoutput=1
\usepackage[a4paper]{geometry}
 
\usepackage[T1]{fontenc} 
\usepackage{comment}

\usepackage{jheppub,hyperref,float,array,adjustbox,mathtools,physics,xcolor}
\usepackage{lipsum}
\usepackage{booktabs}
\usepackage{pdflscape}
\usepackage{geometry}
\usepackage{fancyhdr}
\usepackage{changepage}
\usepackage[english]{babel}

\usepackage{xcolor,tikz,pgfplots,amsmath,amsfonts}
\usepackage{graphicx}
\usepackage[mathscr]{euscript}
\usetikzlibrary{matrix,calc,positioning,decorations.markings,decorations.pathmorphing,decorations.pathreplacing}
\usetikzlibrary{arrows,cd}
\usetikzlibrary{shapes.misc}
\usetikzlibrary {shapes.geometric}
\usetikzlibrary {shapes.multipart}
\usetikzlibrary{decorations.markings}
\usetikzlibrary{backgrounds}
\usetikzlibrary{shapes.gates.logic.US}
\usetikzlibrary{circuits.ee.IEC}
\usepackage{bbding}

\definecolor{terradisiena}{RGB}{233,116,81}
\definecolor{strisciadipietro}{RGB}{229,204,255}
\definecolor{verdepetrolio}{RGB}{33,100,119}

\usepackage{tikz}
\usetikzlibrary{automata,positioning,calc}
\usetikzlibrary{decorations.markings}
\tikzset{->-/.style={decoration={markings, mark=at position #1 with {\arrow{>}}},postaction={decorate}}}
\tikzset{-<-/.style={decoration={markings, mark=at position #1 with {\arrow{<}}},postaction={decorate}}}
\tikzset{auto shift/.style={auto=right,->, to path={ let \p1=(\tikztostart),\p2=(\tikztotarget), \n1={atan2(\y2-\y1,\x2-\x1)},\n2={\n1+180} in ($(\tikztostart.{\n1})!1mm!270:(\tikztotarget.{\n2})$) -- ($(\tikztotarget.{\n2})!1mm!90:(\tikztostart.{\n1})$) \tikztonodes}}}

\usetikzlibrary {shapes.geometric}

\definecolor{USPcol}{rgb}{0.94, 0.73, 0.80}
\definecolor{SUcol}{rgb}{0.894, 0.902, 0.976}
\definecolor{SOcol}{rgb}{0.5, 0.85, 0.8}

\title{
\begin{center}
$N^{3/2}$ Scaling from $3d$ $\mathcal{N}=2$ Dualities: \\
\Large{an Alternative Approach to  Chiral Quivers}
\end{center}
}

\author[a]{Antonio Amariti,}	
\author[a,b]{Giulia Lanzetti}

\affiliation[a]{INFN, Sezione di Milano, Via Celoria 16, I-20133 Milano, Italy}
\affiliation[b]{DiSAT, Universit`a degli Studi dell’Insubria, via Valleggio 11,
Como, 22100, Italy}

\emailAdd{antonio.amariti@mi.infn.it}
\emailAdd{glanzetti@uninsubria.it}

\abstract{
We investigate families of 3d $\mathcal{N}=2$ chiral quiver gauge theories conjectured to be dual to M2-branes probing toric SE$_7$ singularities. Geometrically, these families correspond to toric diagrams without internal points. At the field theory level, the models are constructed via an un-higgsing procedure applied to non-chiral quivers. While the moduli space of these theories was shown to match M-theory expectations, determining the $N^{3/2}$ scaling of the free energy remained an open problem for over a decade, with positive results emerging only very recently.
In this work, we address this challenge by reformulating the three-sphere partition function as a hyperbolic hypergeometric integral. Using exact integral identities, we show that the free energy reduces precisely to that of non-chiral quivers with chiral flavors, for which the $N^{3/2}$ scaling is already established. Physically, this mathematical identity corresponds to the equivalence of three-sphere partition functions under a generalization of Giveon-Kutasov duality to chiral quivers. Our results thus provide a large $N$ duality between the chiral quivers and non-chiral quivers with chiral flavors, confirming the $N^{3/2}$ scaling for the chiral quivers under study.
}

\begin{document}
\maketitle
\flushbottom
\allowdisplaybreaks

\section{Introduction}
\label{sec:intro}

A  strong test of the holographic correspondence regards the analysis of the scaling of the degrees of freedom of a CFT with respect of its gravitational counterpart. When considering 10d Type II and 11d M-theory backgrounds on AdS$_d \times \mathcal{M}_p$, with a compact $\mathcal{M}_p$
manifold  the relevant quantity that accounts for such scaling on the gravitational side is associated to the volume of the compact manifold. A generic discussion for extracting the volume of Sasaki-Einstein spaces has been worked out in seminal papers \cite{Martelli:2005tp,Martelli:2006yb}. 
The field theoretical quantity then depends on the dimensionality of the CFT. It is either related to a trace anomaly (such as the coefficient of the Euler density in four dimensions) or to the partition function on a maximally symmetric compact space (e.g., the free energy on  $S^3$ in three dimensions). 
In the specific case of 3d CFT originating from $N$ M2 branes probing the tip of a CY$_4$ fourfold over a SE$_7$ basis  the expected scaling of the free energy is of order $N^{3/2}$ \cite{Klebanov:1996un}.
Finding this specific scaling from field theory calculations of the $S^3$  free energy has historically been a non-trivial task and an open problem for over a decade.

A significant breakthrough was due to \cite{Drukker:2010nc} where
the first quantitative microscopic explanation of the $N^{3/2}$ scaling was 
furnished in the case of the ABJM theory. Subsequently the analysis presented in \cite{Herzog:2010hf}  for  $\mathcal{N}>2$ supersymmetry  provided a strong non-perturbative checks of the holographic correspondence in many other models. The analysis of the
$\mathcal{N}=2$  cases became then possible thanks to the powerful localization results of \cite{Jafferis:2010un,Hama:2010av}. In that context, the analysis was pursued in  \cite{Jafferis:2011zi} (see also \cite{Martelli:2011qj,Cheon:2011vi}), but the general results applied only to a subset of the myriad dual quivers proposed in the literature \cite{Hanany:2008cd,Hanany:2008fj,Martelli:2008si,Amariti:2009rb,Davey:2009sr,Franco:2009sp,Hanany:2009vx,Ueda:2008hx,Aganagic:2009zk,Davey:2009qx,Benishti:2009ky,Benini:2009qs,Benishti:2010jn,Klebanov:2010tj,Davey:2011mz}. While these initial results did not rule out the other cases, they posed severe constraints on their existence, despite the fact that their moduli spaces often correctly reproduced the expectations from the holographic correspondence.

The zoology of the 3d models conjectured in the literature to be dual to stacks of $N$ M2 branes probing a CY$_4$ fourfold can be divided in two main classes.
The first class of models is obtained by considering a 4d quiver describing a stack of D3 branes at a toric CY$_3$ singularity. The 3d theory is then constructed by dimensionally reducing the matter content for the multiplets in such  quivers and
adding an integer CS term $k$ to each $(S)U(N)$ gauge node, such that the sum of all the CS vanishes. 
A generic algorithm to work out the toric diagram for models of this class was constructed in \cite{Hanany:2008cd}.
Observe that the recipe just spelled out does not necessarily require a 4d SCFT, and also quivers with self intersecting zig-zag paths can be the starting point for 3d  SCFT describing M2 branes probing $CY_4$.
As noted above, a crucial quantity that must match across the duality is the large $N$ scaling of $F_{S^3}$ and its relation to the volume of the SE$_7$  base. Holography consistently fixes this scaling to $N^{3/2}$. However, previous studies  \cite{Jafferis:2011zi,Martelli:2011qj,Cheon:2011vi,Amariti:2011uw,Amariti:2012tj} observed this specific scaling only for a subset of the quivers within this first class.
As a net result, only the "non-chiral" quivers—where each pair of nodes is connected by an equal amount of bifundamentals and anti-bifundamentals—exhibited the required scaling. If one considers chiral quivers, the $N^{3/2}$ scaling was not observed numerically, and the general analytic rules developed in  \cite{Jafferis:2011zi}  did not apply. The scaling behavior of these chiral models thus remained a significant open question. 
Nevertheless a second class of quivers, proposed in \cite{Benini:2009qs} (see also \cite{Cremonesi:2010ae}), can describe some singularities that the first class cannot realize  (\emph{e.g.} $Q^{111}$).
This second class allows for flavor nodes, where the chirality issue discussed above does not apply (a gauge node can connect to a flavor node via either a fundamental or an anti-fundamental representation). It was shown in \cite{Jafferis:2011zi} that the free energy of these quivers \emph{does} scale as $N^{3/2}$, provided that the bifundamentals come in conjugated pairs.

In a very recent paper \cite{Hosseini:2025jxb}, the question of the scaling of the free energy for some of the chiral models of the first type has been successfully reconsidered. 
Using a stable numerical continuation method that directly solves the saddle-point equations the free energy for a pair of models in this class has been recently evaluated, correctly reproducing the expected results from holography.
This remarkable result applies only for some models in this class, indeed it applies for $Q^{111}$ and for $D_3$, failing for others such as $M^{111}$ and $Q^{222}$.

In this paper we interpret this result by leveraging exact results from the integral identities derived from 3d $\mathcal{N}=2$ Giveon-Kutasov (GK) dualities \cite{Giveon:2008zn}. Even if the duality was originally proposed for 3d CS SQCD, it was then extended to 3d CS-matter quivers in  \cite{Aharony:2008gk,Amariti:2009rb,Kapustin:2010xq,Kapustin:2010mh,Benini:2011mf,Closset:2012eq}.
Here we focus on a specific family of chiral quivers obtained via a systematic "un-higgsing" procedure from known non-chiral quivers 
\cite{Benishti:2009ky,Taki:2009wf,Aganagic:2009zk,Phukon:2011hp}.
 We apply a GK duality transformation to the nodes generated by this un-higgsing (which are $U(N)$ nodes with balanced matter content). Restricting the CS levels to $k=\pm 1$, we observe a  large $N$ duality: the dual quiver effectively becomes a non-chiral quiver with chiral flavor nodes - a class where the $N^{3/2}$ scaling is well established.

At the level of the large $N$ free energy, the fugacities of the newly generated $U(1)$ nodes behave as gauged baryonic symmetries. The leading-order free energy becomes independent of these flat directions \cite{Jafferis:2011zi}. In this sense, we demonstrate that at large $N$, the chiral quivers under inspection are dual to the non-chiral quivers with chiral flavor content. This duality is independently confirmed at the level of the moduli space, as both model types share the same toric diagram.

Our findings analytically support the recent numerical calculations carried out in \cite{Hosseini:2025jxb}. Furthermore, our approach explains the limitations observed in that work: models associated with toric diagrams that have internal points (like $M^{111}$ and $Q^{222}$) cannot be obtained by un-higgsing any standard non-chiral model, and thus their scaling behavior cannot be mapped via GK duality to a known scaling result. This aligns  with the observation in \cite{Hosseini:2025jxb}  that the $N^{3/2}$ scaling was absent for those specific cases.

The paper is organized as follows. In Section \ref{infinite} we discuss the family of chiral quivers considered in the rest of the paper. They consists of quivers with bifundamental matter fields connecting the various gauge nodes. Such quivers are build from 
non-chiral 4d toric quivers, denoted as grandparents below. The effective 4d parents are then obtained by an un-higgsing 
procedure, where a bifundamental is traded in favour of an additional gauge node connected with the rest of the quiver through 
a pair of bifundamental. Such process makes the final quiver chiral and in 4d it is associated to a limiting case of Seiberg duality, i.e. the new gauge nodes has a quantum deformed moduli space and the original quiver is recovered on a branch of the moduli space. While this procedure does not give rise to a 4d SCFT (indeed there are self-intersecting zig/zag paths), the 3d quiver obtained from such parents, if opportunely decorated by a consistent assignation of CS levels, can in principle give rise to a toric SCFT, at least a the level of the moduli space. We provide a constructing algorithm in order to obtain the 3d chiral quiver
 starting from the 3d toric diagram, identifying the 4d grand-parent model, the un-higgsing and the assignation of CS levels.
 We provide also an example that plays a crucial role in the analysis below, consisting of the chiral quiver associated to the
 $Q^{111}$ singularity.
 Then in Section \ref{GKsec} we discuss an exact identity that holds at the level of the squashed three sphere partition function and that represents at physical level the manifestation of the Giveon-Kutasov (a.k.a. ABJ duality \cite{Aharony:2008gk}  when referring to quiver gauge theories) duality from the localization perspective. Here we restrict to the case of the round sphere and for this reason we translate the identity in a language that is more useful for the large $N$ evaluation of the three sphere free energy.
 Then in section  \ref{csecex} we apply the identity to various examples. While some of the examples coincide with the one discussed in \cite{Hosseini:2025jxb}, some other are new. After applying the duality rules we end up with quivers with a non-chiral matter content among the nodes of rank $N$, while the chiral matter content is related to bifundamental fields with one index charged under an $U(1)$ gauge symmetry. We observe that at large $N$ the free energy can be computed using the approach of \cite{Jafferis:2011zi} and in this way we predict the off-shell behavior in terms of the charges of the original quiver as well, thanks to the duality dictionary. We complete the analysis connecting our results with the prediction from the geometry and from volume minimization.
Then in section \ref{secgen} we provide various generalization of our work, discussing cases with higher CS levels,  with more gauge groups, and commenting on the case of vanishing CS level for the un-higgsed node and on the relation with tensor deconfinement. We further provide a discussion for a chiral case with five gauge groups. This case is interesting also for other reasons, associated to the existence of a quartic formula for the geometric free energy. We provide the quartic formula in this case and comment on some novelties in the expression with respect to existing literature.
Then in section  \ref{discussion} we provide some conclusive remarks.\
We also added two appendices. In appendix  \ref{toricdiagram} we give some technical details for the computation of the toric diagram,  useful in the formulation of the inverse algorithm discussed in the body of the paper. In appendix 
\ref{onegroup} we provide a discussion for the un-higgsing and the relation with GK duality for quivers with non-chiral
field content.

\section{An infinite family of chiral quivers}
\label{infinite}

In this section we describe the infinite family of chiral quivers that we consider in the rest of the paper and 
that we claim to have an $N^{3/2}$ scaling of the free energy.
The models are chiral quivers build  from non-chiral quivers through an un-higgsing procedure. 
Here we shown how to obtain these models starting from a 3d toric diagram, i.e. by exploiting an \emph{inverse}  algorithm.

For this reason we start by considering a 3d toric diagram and we intersect it with a plane living on a 
two dimensional integer lattice.
We then  project the 3d toric diagram on this 2d lattice,  obtaining a 
2d toric diagram. Observe that different intersecting lattices correspond to different
2d toric diagrams.
We define a projection to be  \emph{consistent} if all the points of the 3d toric
diagram lies on the lattice
(i.e. their coordinate on the sub-lattice have to be integer). Otherwise the projection is
not consistent.

Furthermore, before projecting the 3d toric diagram, we can make an $SL(3,\mathbb{Z})$ transformation
on it. In such a way all the 2d sub-lattices we project on correspond 
to the planes $(x,y)$, $(x,z)$ or $(y,z)$.

After we project the 3d toric diagram on a 2d one 
we can exploit the inverse algorithm of \cite{Hanany:2005ss}. This gives rise to a 
4d consistent \emph{parent} quiver gauge theory associated to the 2d
toric diagram.
Observe that the 2d toric diagram associated to the four dimensional  
theory has vertices with multiplicity one and 
points on the edges with multiplicity given by the combinatorics.
There can also be internal points which multiplicity depends from the Seiberg 
dual phase we are considering. 
Here we restrict to general our attention to 2d toric diagrams without any internal point,
because they are the starting point to build 3d chiral quivers with the $N^{3/2}$ scaling.
However the algorithm that we are building works, at the level of the moduli space, also for toric diagrams with internal points.
If all the vertices of the projected 2d toric diagram obtained so far have multiplicity 1, then we assign the CS levels
to the $(S)U(M)$ gauge groups, such that their sum is vanishing,  and we compute the toric diagram of the three dimensional theory associated to the quiver that we found, using the rules discussed in Appendix \ref{toricdiagram}.
 If there exist an assignation of levels such that we reproduce the original toric diagram, we have found one toric theory directly from the toric diagram.

Anyway, the multiplicity of the points of the two dimensional projected diagram 
corresponds to the number of vertices lying
on the  direction orthogonal to the projection 2d lattice and in many cases
these vertices do not have multiplicity 1 
An algorithm that generates a CS three dimensional quiver from the toric data
must reproduce this higher multiplicity  from the four dimensional gauge theory.

In the next subsection we explain a procedure to construct  4d 
quiver gauge theories where the vertices have multiplicity higher than
one.  We will observe that, even if the 4d parent theory becomes inconsistent,  the 3d CS quiver can still give origin to a consistent SCFT. For this reason the  3d  theories discussed here do not 
have in general a four dimensional consistent toric parent.

\subsection{3d theories without a 4d parent}

As shown in \cite{Taki:2009wf} the knowledge of the PM is enough to compute the toric 
diagram of a 3d CS matter quiver  theory with a 4d consistent toric parent.
Here we introduce the algorithm that we will use in the rest of the paper.
After we project the 3d toric diagram on the 2d lattice, we obtain the toric diagram of candidate 4d parent. Usually this diagram has some external point with multiplicity higher than 1, i.e. it is associated to an inconsistent model.

In order to obtain such an inconsistent  model 
we start by considering a consistent 2d toric diagram, i.e. without further multiplicities for the external points. 
Then  we study the increasing of the multiplicity of the external points by acting on this consistent four dimensional theory, that we refer to as the gran-parent theory.
Crucially we increase the multiplicity of the vertices by increasing
the number of perfect matchings associated to the 
vertices of the 2d diagram. 
This is done by splitting one of the fields in the perfect matching
associated to a  vertex.  This procedure is called un-higgsing \cite{Feng:2002fv}, because the inverse procedure coincides with giving a vev to a bifundamental field charged under the gauge groups.
For example if an external point is associated to a perfect matching 
$X_{ij} \dots X_{\alpha \beta} \dots X_{kl}$ we can split the field $X_{\alpha \beta}$ such that it becomes $X_{\alpha A} X_{A\beta}$, where $A$ is the index of a new gauge group.
The superpotential remains toric in the combination   $X_{\alpha A} X_{A\beta}$,
and there is an external point which has now higher multiplicity. Actually all the
points associated with a perfect matching including $X_{\alpha \beta}$ are doubled.

\begin{figure}[ht] \begin{center}
\includegraphics[width=12cm]{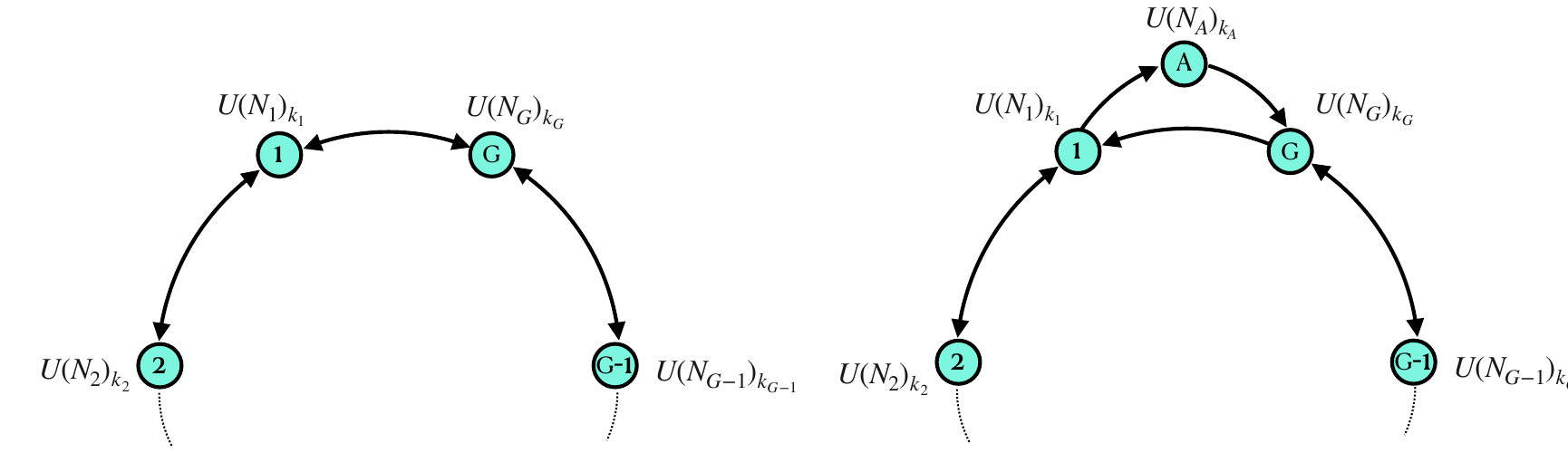}
\caption{Quiver and CS levels for the 3d model descending from a 4d consistent grand-parent theory and its un-higgsed 
version descending from a 4d inconsistent (not a SCFT) parent theory. }\label{tilingprimaedopo}
\end{center} \end{figure}

This procedure makes the 4d parent theory inconsitent, because there are fields with zero R charge (or equivalently there is a gauge group with $N_f=N_c$, which cannot be described classically).
Anyway the CS matter theory in 3d associated with this quiver does still make sense and 
if there is an assignation of CS levels such that this 2d toric diagram splits in the original 3d
one, we have just built the quiver gauge theory directly from the toric data.
The case in which just one field is involved in this un-higgsing is represented pictorially in Figure
 \ref{tilingprimaedopo} where the new group is denoted as $A$ in the Figure.

While the consistent grand-parent theory has $G$ gauge groups, the non-consistent one, the parent theory in this case  has $G+1$ groups. 
Starting from the original CS levels $k_1,\dots k_G$, satisfying 
$\sum k_i=0$, we assign CS level $k_A$ to the $G+1$-th group and modify
the constraint to $\sum k_i=-k_A$. This is achieved by keeping $k_1, \dots, k_{G-1}$ fixed while adjusting the level of the last group to
$k_G = -k_1-\dots-k_{G-1}-k_A$. \\
This CS levels assignment is automatically implemented in the algorithm reviewed in Appendix \ref{toricdiagram} 
for the computation of the toric diagram of the un-higgsed theory. It leaves the CS level fluxes as in the initial case, while the
new fields arising from the un-higgsing have a CS flux assigned as in Figure \ref{tilingprimaedopo}.\\ 
The toric diagram remains unchanged for the PMs not including $A$, whereas those containing $A$ have a split $(0,k_A)$ in the CS direction.
Indeed, each column of the PM matrix corresponds to a point in the toric diagram, with the same ordering maintained in $G_t$. If a column in $P$ is doubled because of the un-higgsing $X_{ij} \rightarrow X_{iA} X_{Aj}$, then one of the new points remains identical to the original, while 
the other shifts by an additional $k_A$ contribution. All the points 
where $X_{ij}$ is absent are unchanged.

\begin{figure}[ht]
\centering
\includegraphics[width=7cm]{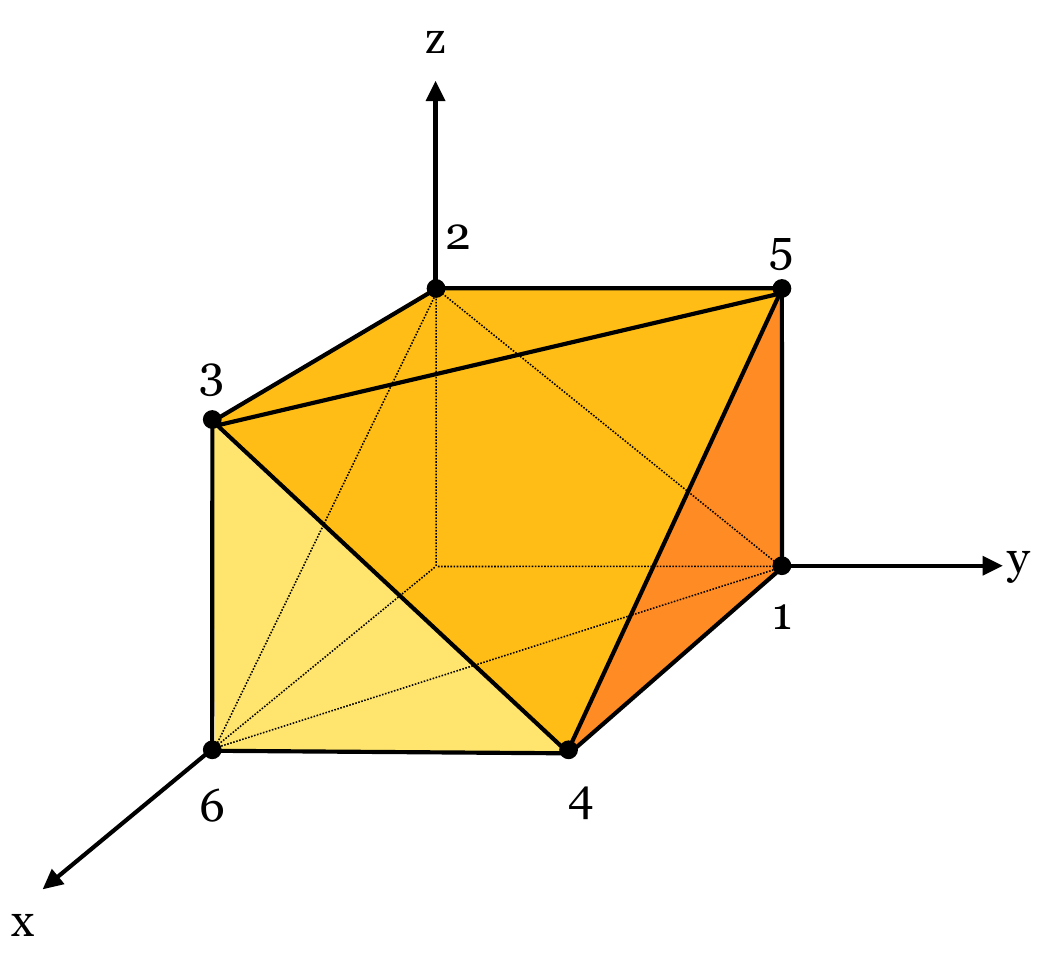}
\caption{Three dimensional toric diagram for $Q^{111}$, built here from the superpotential (\ref{figu}).}
\label{ABJMmod}
\end{figure}

\subsection*{Example: $Q^{111}$}

\begin{figure}[H]
\centering
\includegraphics[width=10cm]{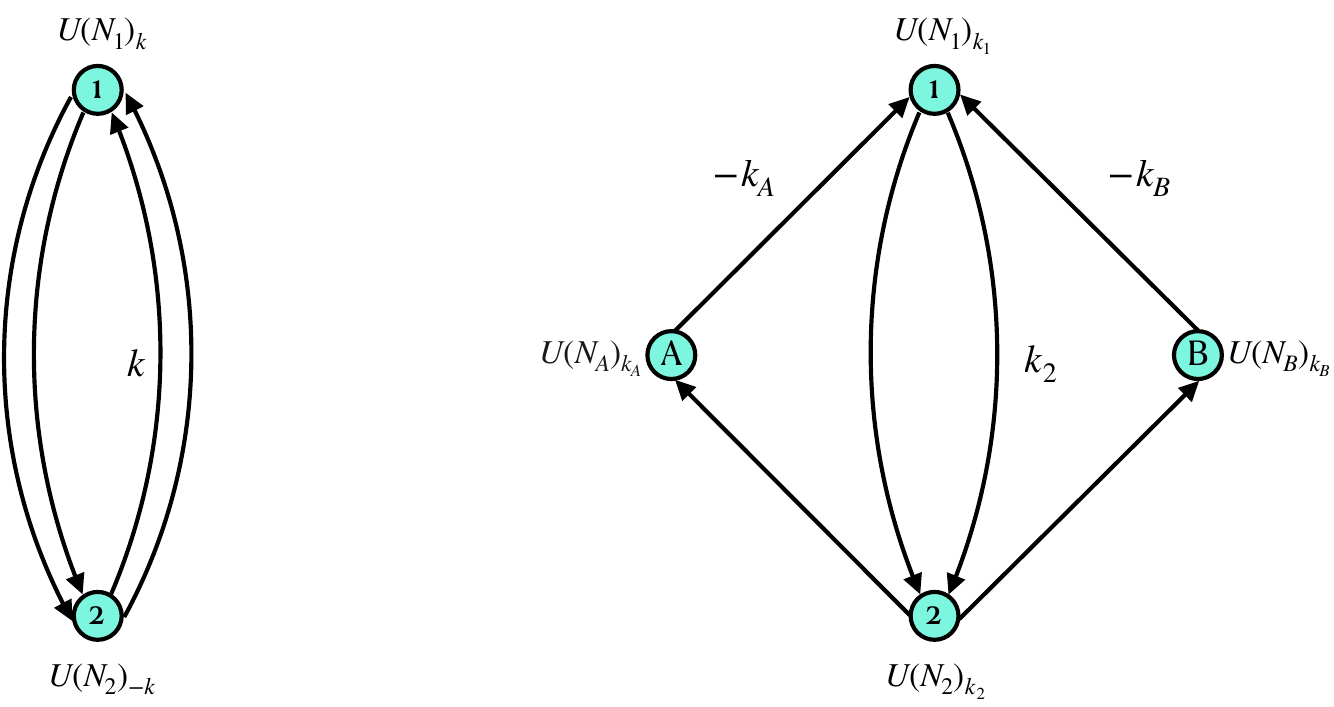}
\caption{ABJM quiver and the un-higgsed quiver corresponding to $Q^{111}$ with the CS levels $k_A =-k_B=-1$ and $k_1=k_2=0$.}
\label{newquivdue}
\end{figure}

At this point of the discussion we can study an example of a theory without a consistent  four dimensional parent.
We consider the toric diagram for $Q^{111}$ given in Figure \ref{ABJMmod}. 
By projecting it on the $(X,Y)$ 2D lattice we obtain the 
toric diagram of the 4d conifold, but where two of the external points have double multiplicity.\\
The two external points that we modify are associated to the third and the fourth column of the perfect matching matrix in (\ref{PMconi}). By doubling these points, we double two of the perfect matchings. As explained above the doubling of the perfect matchings is done by splitting some of the fields in the superpotential and by generating some new gauge groups. In this case the fields are split as
\begin{equation}
	X_{21}^{(1)} \rightarrow X_{2A} X_{A1}
	\quad , \quad
	X_{21}^{(2)} \rightarrow X_{2B} X_{B1}
\end{equation}
The moduli space of 
this theory with four gauge groups 
gets quantum correction because the groups $SU(N_A)$ and 
$SU(N_B)$ have $N_F=N_C$ if 
$N_A=N_B=N_1=N_2=N$.
It is not the dual field theory 
describing N D3 branes at 
some toric singularity in 4d.
The superpotential of this theory is
\begin{equation} \label{figu}
	W = X_{12}^{(1)} X_{2A} X_{A1} X_{12}^{(2)} X_{2B} X_{B1} - X_{12}^{(1)} X_{2B} X_{B1} X_{12}^{(2)} X_{2A} X_{A1}
\end{equation}
Anyway this theory can be consistent in three dimensions and an opportune choice of the CS levels leads indeed to the 
toric diagram in Figure \ref{ABJMmod}.
The splitting changes the quiver and we can define the CS
levels as in Figure \ref{newquivdue}.
The adjacency matrix and the perfect matching matrix become
\begin{equation}
d=
\left(
\begin{array}{c||cccccc}
&X_{A1}&X_{2A}&X_{B1}&X_{2B}&X_{12}^{(1)}&X_{12}^{(2)}\\
\hline
\hline
N_1&-1&0&-1&0&1&1\\
N_2&0&1&0&1&-1&-1\\
N_A&1&-1&0&0&0&0\\
N_B&0&0&1&-1&0&0\\
\end{array}
\right) ,\quad
P  =
\left(
\begin{array}{c||cc|cc|cc}
X_{A1}&1&0&0&0&0&0\\
X_{2A}&0&1&0&0&0&0\\
X_{B1}&0&0&1&0&0&0\\
X_{2B}&0&0&0&1&0&0\\
X_{12}^{(1)}&0&0&0&0&1&0\\
X_{12}^{(2)}&0&0&0&0&0&1\\
\hline
\hline
\text{CS}&-k_A&0&-k_B&0&0&k_2
\end{array}
\right)  
\end{equation}
The last row is given by the CS row of the perfect matching matrix.
By choosing $(k_1,k_2,k_A,k_B) = (0,0,-1,1)$
we obtain the desired toric diagram
\begin{equation}
\label{torcDQ111}
G_t^{(3D)}=\left(
\begin{array}{cccccc}
1 & 1& 0 & 0 & 0 & 1 \\
 0 & 0 & 1 & 1 & 0 & 1 \\
 0 & 1 & 0 & -1& 0 & 0 \\
 1 & 1 & 1 & 1 & 1 & 1 \\
\end{array}
\right)
\end{equation}

and as anticipated above, two points of the four dimensional toric diagram are doubled.
Notice that there are other assignments of CS levels that are toric dual to the one discussed here (see e.g. \cite{Franco:2009sp,Davey:2009qx}).

\section{GK duality in a chiral quiver}
\label{GKsec}
In this section we discuss how to reduce an $U(N)_k$ integral into a $U(|k|)_{-k}$ integral in the case of a chiral quiver with  only $U(N)$ gauge factors.
This reduction at the level of the integrals corresponds to a physical duality that generalizes the GK duality \cite{Giveon:2008zn}, originally worked out for SQCD with a single $U(N)_k$ gauge factor from the brane perspective and then obtained at field theoretical level from a real mass flow from Aharony duality \cite{Aharony:1997gp} (where the CS level is vanishing).
The generalization of  GK duality to quiver gauge theories has already been discussed in various papers in the literature, originally for quivers with a non-chiral matter content \cite{Kapustin:2010mh}, and then for case with chiral matter content \cite{Benini:2011mf}. 

In the non chiral, when considering the ABJM quiver,  the duality has been originally extracted from the brane setup in \cite{Aharony:2008gk}. It corresponds to an Hanany-Witten transition in the case of elliptic models with D3 brane wrapped on a circle and intersecting fivebranes corresponding to bound states of NS and D5 branes.
In this setup, GK duality not only modifies the gauge group undergoing the Hanany–Witten transition, but also affects its nearest-neighbor nodes by shifting their Chern–Simons levels by the level of the node that undergoes the duality.
However, if we consider a chiral model, the rules of the duality require a more sophisticated analysis as done in \cite{Benini:2011mf}, although the latter did not focus on the effect of the duality on the abelian gauge factors.

Here we reconsider the problem restricting our interest to the three sphere partition function and for this reason we focus on the exact integral identity that appeared in the mathematical literature and that is interpreted as the GK duality when one restricts to $U(N)_k$ SQCD.
The relations is  \cite{vandeB:2007}
\begin{equation}
\label{GKrule}
Z_{U(N)_k}^{F \square,F \overline \square}(\vec \mu;\vec \nu;\lambda)
=
\prod_{a,b=1}^F \Gamma_h(\mu_a + \nu_b) e^{-\frac{i \pi}{2} \phi}
Z_{U(F-N+k)_{-k}}^{F \square,F \overline \square}(\omega-\vec \nu;\omega-\vec \mu;-\lambda)
\end{equation}
where we fixed $k>0$ and where each integral is defined as
\begin{equation}
\label{rules}
Z_{U(N)_k}^{F \square,F \overline \square}(\vec \mu;\vec \nu;\lambda)
=
\frac{1}{N! }
\int \prod_{i=1}^N d \sigma_i e^{ i \pi k \sigma_i^2-  i \pi  \lambda \sigma_i}
\frac{\prod_{a=1}^F \prod_{i=1}^{N} \Gamma_h(\mu_a-\sigma_i,\nu_a + \sigma_i)}{\prod_{1\leq i <j \leq N} \Gamma_h(\pm (\sigma_i -\sigma_j))}
\end{equation}
and the phase is given by
\begin{eqnarray}
\label{phase}
\phi&=&\left(-
k\left(\!\sum_{r=1}^{F} \mu_r^2 \!+\!\sum_{s=1}^{F} \nu_r^2\!\right)\!+\! 
k(k-2\tilde N)\omega^2+\frac{1}{2}\lambda^2
\!+\!2 k\left(\!\sum_{r=1}^{F} \mu_r\!+\!\sum_{s=1}^{F} \nu_s\!\right)\omega\!\right)
\nonumber \\
&+&
\left(\lambda\left(\sum_{r=1}^{F} \mu_r-\sum_{s=1}^{F}\nu_s\right)+
\frac{1}{2}(2\tilde N\omega -\sum_{r=1}^{F}\mu_r-\sum_{s=1}^{F} \nu_s)^2\right)
\end{eqnarray}
An analog relation can be worked out for $k<0$  by sending $k\rightarrow -k$ in the phase (this is a parity transformation).\\
In the formula above we have used the conventions adopted in the mathematical literature, where the one loop determinants are expressed in terms of the hyperbolic Gamma function. It corresponds to the regularized partition function\footnote{We refer the reader to \cite{Ruijsenaars1997} for the regularization procedure.} 
on the squashed three sphere and it reduces to the round case by setting the squashing parameter to $1$.
Before its regularization  the (squashed) three sphere partition function is a divergent quantity, given by the formula 
\begin{equation}
\Gamma_h(z;\omega_1,\omega_2)=\prod_{n_1,n_2 \geq 0}^{\infty}	\frac{(n_1+1)\omega_1+(n_2+1)\omega_2- z}
	{n_1 \omega_1+n_2 \omega_2+z} \, ,
\end{equation}
where $\omega_1 = i b$ and $\omega_2=i/b$ are the squash parameters.

Observe that in the formulas above we adopted the shortcut $\Gamma_h(z)$ instead of 
$\Gamma_h(z;\omega_1;\omega_2) $ for simplicity. When $b\rightarrow 1$ the squashed sphere becomes the round sphere and indeed 
 we have $\Gamma_h(z;i;i) = e^{\ell(1+i z)}$, where the function $\ell(z)$ is the one-loop determinant originally found in \cite{Jafferis:2010un} and it is the function used in general to study the scaling of the free energy at large $N$ and  F-maximization.
  In the following we restrict our attention to the case of the round three sphere, 
  even if the discussion holds more generally on the squashed three sphere for the examples considered below.
 
\begin{figure}
\begin{center}
  \includegraphics[width=7cm]{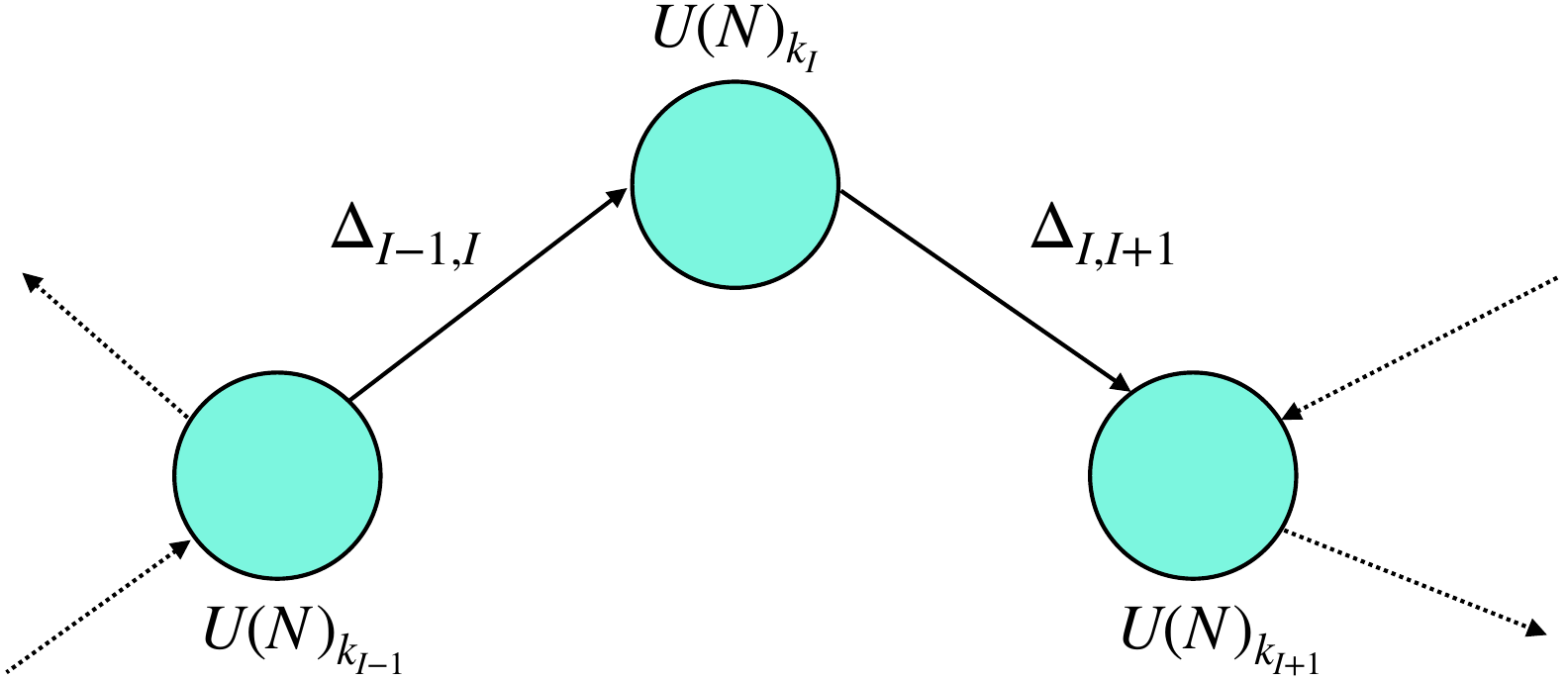}
  \end{center}
  \caption{In this figure we represent three $U(N)$  gauge nodes embedded in a generic chiral quiver. This is the typical situation that we encounter in this paper. Each node has a different CS level, $k_{I-1}$, $k_{I}$ and $k_{I+1}$ in the figure. When $k_I\neq 0$ we can dualize the node at the level of the three sphere partition function using an  exact mathematical identity, corresponding to the GK duality for a chiral quiver. In the figure we also assigned to the two bifundamentals connecting the three gauge nodes the relative R-charge. At this level of the discussion we did not specify any possible further constraints between such charges, that depend on the detail of the model under inspection. 
    }
    \label{Fig:GK}
\end{figure}

 We then consider a chiral quiver and apply the rule (\ref{GKrule}) to the $I$-th node.
 The transformation is local in the quiver and it involves only the three nodes represented in Figure \ref{Fig:GK}. The three sphere partition function for the quiver corresponds to an integral over the variables $\sigma_{i}^{(I)}$ where the label $i$ runs over the Cartan of each $U(N)$ factor and $I$ runs over the number of gauge nodes.
The term that is involved in the transformation (\ref{GKrule}) at the level of the partition function is then
\begin{equation}
\label{localint}
\int \prod_{i=1}^N d \sigma_i^{(I)} e^{i \pi k_I \sigma_i^{(I)2}+2 \pi \Delta_{m_I} \sigma_i^{(I)}}
\frac{\prod_{i,j=1}^N 
e^{\ell(1-\Delta_{I-1,I}+i(\sigma_i^{(I-1)}-\sigma_j^{(I)}))
+
\ell(1-\Delta_{I,I+1}+i(\sigma_i^{(I)}-\sigma_j^{(I+1)}))
}}{
\prod_{1\leq i <j \leq N} 
\left(2 \sinh (\pi(\sigma_i^{(I)} -\sigma_j^{(I)}))\right)^{-2}} \, .
 \end{equation}
 Comparing with the generic formula (\ref{rules}) we have
\begin{equation}
\label{parmasses}
\left\{
\begin{array}{ccl}
\mu_a&=& i \Delta_{I-1,I}+ \sigma_{a}^{(I-1)} \\
 \nu_a&=&  i \Delta_{I,I+1} -\sigma_{a}^{(I+1)}\\
 \lambda_I&=&2 i  \Delta_{m_I} 
 \end{array}
 \right. \quad a=1,\dots,N\,.
 \end{equation}
 At this point we can plug in the masses parameterized as in \eqref{parmasses} in the relation (\ref{GKrule}). The integral (\ref{localint}) becomes  
  \begin{eqnarray}
  &&
 e^{-\frac{i \pi \phi }{2} } 
\prod_{i,j=1}^{N}   e^{\ell(1-\Delta_{I-1,I}-\Delta_{I,I+1}+ i(\sigma_i^{(I-1)}-\sigma_j^{(I+1)}))}
  \\
 &&
  \int  \prod_{a=1}^{|k_I |} d \sigma_a^{(I)} e^{-i \pi k_I {\sigma_a^{(I)}}^2-2 \pi \Delta_{m_I} \sigma_a^{(I)}}
\prod_{j=1}^N 
e^{\ell(\Delta_{I-1,I}-i(\sigma_j^{(I-1)}-\sigma_a^{(I)}))
+
\ell(\Delta_{I,I+1}-i(\sigma_a^{(I)}-\sigma_j^{(I+1)}))}\nonumber
 \end{eqnarray}
 where the phase can be computed from \eqref{phase} and it is 
  \begin{eqnarray}
  	\phi&=&
  	-k_I \sum _{i=1}^N \left({\sigma_i^{(I-1)}}^2+{\sigma_i^{(I+1)}}^2\right)-\sum _{i=1}^N \sigma_i^{(I-1)} \cdot \sum _{i=1}^N \sigma_i^{(I+1)}
  	\nonumber \\
  	&+&\frac{1}{2} \Big(\sum _{i=1}^N \sigma_i^{(I-1)}\Big)^2+\frac{1}{2} \Big(\sum _{i=1}^N \sigma_i^{(I+1)} \Big)^2
  	\nonumber \\
  	&-&i
  	( (2k_I-N) \Delta _{I-1,I}-N \Delta _{I,I+1}-2 \Delta_{m_I}) \sum _{i=1}^N \sigma_i^{(I-1)} 
  	\nonumber \\
  	&+&i
  	((2 k_I-N) \Delta _{I,I+1}-N \Delta _{I-1,I}+2 \Delta_{m_I}) \sum _{i=1}^N \sigma_i^{(I+1)} \, ,
  \end{eqnarray}
where we omitted the real terms in the phase, because they are irrelevant in our discussion, due to the fact that we focus on the free energy obtained as 
$F = -\log|Z|$.

Observe that the first two lines in this relation signal the presence in the $U(N_{I\pm1})$ gauge nodes of a shift of $k_I/2$ of the CS levels for the $SU(N)$ part, and additional shifts for the $U(1)$ factors. Such shifts at generic $N$ are problematic, because they do not have the right quantization property in order to be associated to gauge invariant CS terms in the underlying lagrangian. 
In the examples below however we will study quivers where only a diagonal $U(1)$ factor is gauged and such terms vanish. 
Furthermore this simplification will affects also the shift in the FI in the last two lines as we will discuss in the various examples.
Gauging only this diagonal $U(1)$ on the three sphere partition function corresponds to impose the constraint $\delta\left(\sum _{i=1}^N(\sigma_i^{(I-1)} -\sigma_i^{(I+1)})\right)$, such that the relevant phase that is not set to zero is
   \begin{eqnarray}
   \label{betterFI}
   \phi=&-&
  	k_I \sum _{i=1}^N {\sigma_i^{(I+1)}}^2
	+
	2 i( \Delta_{m_I} + k \Delta _{I,I+1}) \sum _{i=1}^N \sigma_i^{(I+1)}\nonumber \\&-&
	k_I \sum _{i=1}^N {\sigma_i^{(I-1)}}^2+
2 i( \Delta_{m_I} -k\Delta _{I-1,I})  ) \sum _{i=1}^N \sigma_i^{(I-1)}	
   \end{eqnarray}

\section{Examples: toric duals of ABJM with chiral flavor}
\label{csecex}

In this section we consider chiral gauge theories that are toric dual to models obtained in \cite{Benini:2009qs} by flavoring the quiver of the 4d conifold quiver, with opportune CS levels.
The  models under investigation here correspond to $Q^{111}$ , $D_3$, $\mathcal{C} \times \mathbb{C}$ and the cubic conifold. 
The first two models under investigation are chiral quivers recently studied in \cite{Hosseini:2025jxb}, and our result corroborate the numerical results on the $N^{3/2}$ scaling of the free energy. The other examples have not been studied numerically yet and they correspond to new predictions, that necessitate of a direct saddle point analysis.  
Furthermore the third example $\mathcal{C} \times \mathbb{C}$ shows the generality of our argument, which is valid indeed also for models that have semi-integer CS levels when considering  chiral flavor.
In the following we study the models by first providing the chiral quiver and then, by applying \eqref{GKrule} on the free energy, we show that the models  at order $N^{3/2}$ are dual to the quivers with two gauge groups and chiral flavor defined in \cite{Benini:2009qs}. The free energy of such models have been studied in \cite{Jafferis:2011zi} and by exploiting these results we obtain the free energy for the chiral quivers as well, confirming the $N^{3/2}$ scaling and matching the off-shell results with the volume computation.
A detailed analysis on the fate of the FI terms under duality is crucial for matching the results with the expectation from the geometry.

%
%
%
%
\subsection{The free energy of $D_3$ and GK duality}
\label{subsec:D3}
%
%
%
%

\begin{figure}[H] \begin{center}
\includegraphics[width=6cm]{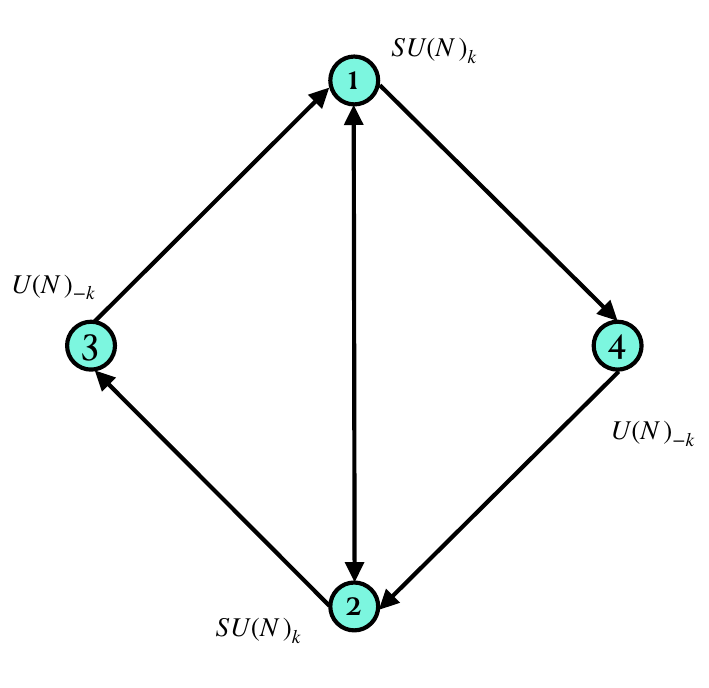}
\caption{Chiral quiver description of the $D_3$ model. 
We did not represent the extra $U(1)_{diag}$ gauging in the picture.}
\label{fig:D3quiv}
\end{center}
 \end{figure}

The first example that we consider is the second case studied in \cite{Hosseini:2025jxb}, corresponding to the quiver in Figure \ref{fig:D3quiv}.
Our main claim is that by acting with the identity \eqref{GKrule} associated to the GK duality, the two models are dual at leading order in $N$, corroborating the validity of the scaling found in \cite{Hosseini:2025jxb}.

Actually before proceeding we need to discuss a crucial point of the models with chiral flavors. As shown in \cite{Jafferis:2011zi}, the relation between F-maximization and Volume minimization requires on the gauge theory side an 
$SU(N)^G \times U(1)_{diag}$ gauge group.
For this reason, in the following we will need to match the computation for the chiral flavor with this gauge group. Then the chiral quiver associated to D$_3$, given in Figure \ref{fig:D3quiv}, has $SU(N_1)_{k_1} \times SU(N_2)_{k_2}$ gauge factors, with in addition a gauged diagonal $U(1)$ factor. On the other hand the other two gauge groups 
are $U(N)$, in order to apply the GK duality and make contact with the large $N$ three sphere free energy of the model with chiral flavor\footnote{See \cite{Benishti:2010jn,Klebanov:2010tj} for detailed analysis of $U(N)$ vs $SU(N)$ gauge nodes.}. 
The superpotential for the chiral quiver is given by
\begin{equation}
W = X_{12} X_{23} X_{31}X_{14} X_{42} X_{21}
-
X_{12} X_{21} X_{14} X_{42} X_{23} X_{31}
\end{equation}

The three sphere partition function is written as  
\begin{eqnarray}
\label{ZD3ele}
Z_{D_3}&=&\frac{1}{N!^4}
\int
\frac{ \prod_{i=1}^{N} \prod_{a=1}^4
d \sigma_{i,a}
e^{i k_a \pi \sigma_{i,a}^2+ 2  \pi \Delta_{m_a} \sigma_{i,a}}}{
\prod_{a=1}^4 \prod_{1\leq i < j \leq N} \left(2 \sinh (\pi(\sigma_{i,a}-\sigma_{j,a}))\right)^{-2}} 
\delta\left({\tiny \sum}_{i=1}^{N} (\sigma_{i,1}-\sigma_{i,2})\right) 
\nonumber \\
&\times &
\prod_{i,j=1}^{N} e^{\ell(1-\Delta_{X_{12}} + i (\sigma_{i,1} - \sigma_{j,2} )) + \ell(1-\Delta_{X_{21}}+i(\sigma_{i,2} - \sigma_{j,1}))
+
\ell(1- \Delta_{X_{23}}+i(\sigma_{i,2} - \sigma_{j,3})) }
 \\
&\times&
e^{\ell(1-\Delta_{X_{31}}+i(\sigma_{i,3} - \sigma_{j,1} ))
+
\ell(1- \Delta_{X_{14}}+i(\sigma_{i,1} - \sigma_{j,4} ))+\ell(1-\Delta_{X_{42}}+i(\sigma_{i,4} - \sigma_{j,2} ))
}
 \nonumber 
\end{eqnarray}
where the CS levels are  assigned as in Figure \ref{fig:D3quiv}.

\begin{figure}[ht] \begin{center}
\includegraphics[width=7cm]{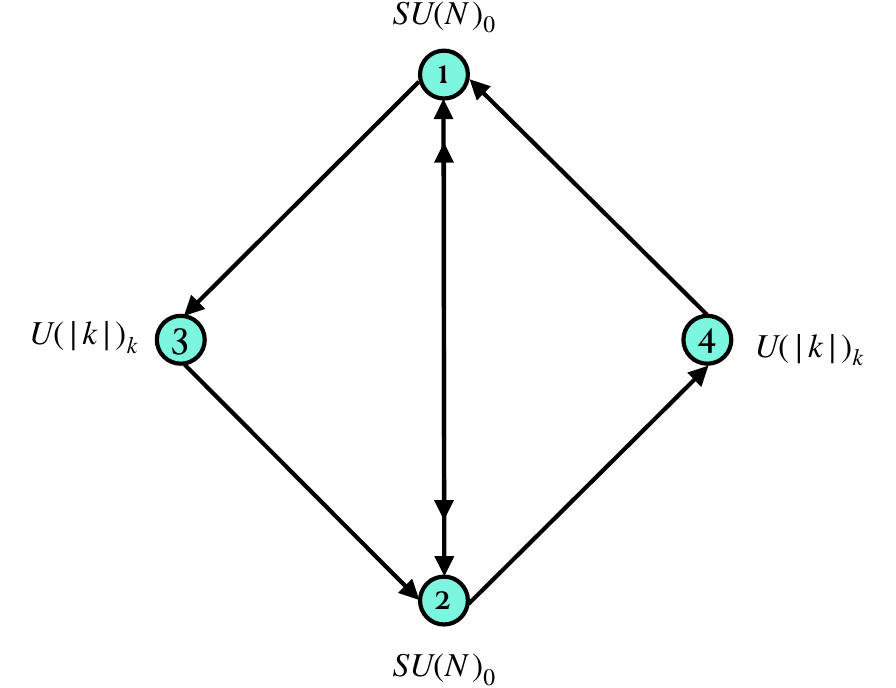}
\caption{Quiver obtained after Giveon-Kutasov duality 
on the chiral quiver describing the $D_3$ model.}
\label{fig:D3dual}
\end{center} \end{figure}

The application of the relation \eqref{GKrule} on the nodes $3$ and $4$ gives rise to the partition function of the model represented by the quiver in Figure \ref{fig:D3dual}.
This phase has superpotential 
\begin{eqnarray}
W = X_{12} Y_{21} Y_{12} X_{21}
-
X_{12} X_{21}  Y_{12}  Y_{21}+Y_{12} Y_{24} Y_{41} +Y_{21} Y_{13}Y_{32}
\end{eqnarray}
In the following we focus on the case $k=1$.
The dual partition function in this case is given by
\begin{eqnarray}\label{ZD3dual}
&&
Z_{D_3}= \frac{1}{N!^2}
\int
\frac{  \prod_{a=1}^2 d \sigma_{a+2} \prod_{i=1}^N 
d \sigma_{i,a}
e^{i \pi (\sigma_{3}^2+\sigma_{4}^2)+2 \pi  \sum_{i=1}^N \sum_{a=1}^2 \tilde \Delta_{m_a} \sigma_{i,a}
-2 \pi \sum_{a=3}^4\Delta_{m_a} \sigma_{a}}}{
\prod_{a=1}^2\prod_{1\leq i < j \leq N} \left(2 \sinh (\pi (\sigma_{i,a}-\sigma_{j,a}))\right)^{-2}} \nonumber \\
&&\delta\left({\tiny \sum}_{i=1}^{N} (\sigma_{i,1}-\sigma_{i,2})\right) 
\prod_{i,j=1}^{N} 
e^{
\ell(1-\Delta_{X_{12}}+i(\sigma_{i,1} - \sigma_{j,2} ))
+
\ell(1-\Delta_{X_{21}}+i(\sigma_{i,2} - \sigma_{j,1} ))
}
 \\
&&
e^{
\ell(1-\Delta_{Y_{12}}+i(\sigma_{i,1} - \sigma_{j,2} ))
+
\ell(1-\Delta_{Y_{21}}+i(\sigma_{i,2} - \sigma_{j,1} ))
+
\ell(1-\Delta_{Y_{32}}+i(\sigma_{i,3} - \sigma_{j,2} ))
}
\nonumber \\
&&e^{
\ell(1-\Delta_{Y_{13}}+i(\sigma_{i,1} - \sigma_{j,3} ))
+
\ell(1-\Delta_{Y_{41}}+i(\sigma_{i,4} - \sigma_{j,1} ))
+
\ell(1-\Delta_{Y_{24}}+i(\sigma_{i,2} - \sigma_{j,4} ))
}
 \nonumber 
\end{eqnarray}
where
\begin{eqnarray}
\label{dicdeltam}
&&
2\tilde \Delta_{m_1} =\Delta_{X_{14}} -\Delta_{X_{31}}+ \Delta_{m_3}  +  \Delta_{m_4} +2\Delta_{m_1}
\nonumber \\
&&
2\tilde \Delta_{m_2} = \Delta_{X_{23}}-\Delta_{X_{42}}+ \Delta_{m_3}  +  \Delta_{m_4} +2\Delta_{m_2}
\end{eqnarray}
In this case the leading $N^{3/2}$ contribution to the free energy has been obtained in \cite{Jafferis:2011zi} and it is
\begin{equation}
\label{D3kleb}
F = \frac{2\pi N^{\frac{3}{2}} }{3} 
\sqrt{
\prod_{Z = \{X,Y\}} (\Delta_{Z_{12}} \!+\!\Delta_{Z_{21}})
\bigg(
(\Delta_{Y_{12}} +\Delta_{Y_{21}})
-
\frac{4 \tilde \Delta_{m}^2  }{\Delta_{Y_{12}} \!+\!\Delta_{Y_{21}}}
\bigg)}
\end{equation}
with $\tilde \Delta_m = \tilde \Delta_{m_1} + \tilde \Delta_{m_2}$.

At large leading order in $N$ the matrix model associated to this theory is identical to the one for the 
quiver with chiral flavor of \cite{Benini:2009qs}.
Indeed at large $N$ the partition function for this last model has additional accidental flat directions, associated to the baryonic symmetries. Such baryonic symmetries are gauged in our case, and they correspond to the two extra $U(1)_{1}$ factors. Once the large $N$ evaluation of the partition function is performed one can further integrated over the two pure CS theories associated to such $U(1)$ without modifying  the large $N$ result.

We can then use the duality dictionary to obtain the free energy of the chiral version of D$_3$ with four gauge groups
of rank $N$. This is obtained by substituting in (\ref{D3kleb}) the relations 
with $\Delta_{Y_{12}} =\Delta_{X_{14}}+\Delta_{X_{42}}$ and $\Delta_{Y_{21}} =\Delta_{X_{23}} +\Delta_{X_{31}} $,
in addition to (\ref{dicdeltam}).
We obtain
\begin{eqnarray}
\label{D3nostro}
F = \frac{4\pi N^{\frac{3}{2}} }{3}
&& \sqrt{
(\Delta _{X_{12}}+\Delta _{X_{21}}) 
(\Delta_{X_{42}}+\Delta_{X_{31}}-\Delta _m) 
(\Delta _{X_{14}}+\Delta _{X_{23}}+\Delta _m) }
\nonumber \\
&&
\sqrt{
(\Delta _{X_{12}}+\Delta _{X_{23}}+\Delta _{X_{31}}) 
(\Delta _{X_{21}}+\Delta _{X_{14}}+\Delta_{X_{42}})
}
\end{eqnarray}
where $\Delta_m = \sum_{a=1}^4 \Delta_{m_a}$.
The $R$-charges at the fixed point in this phase are alle equal to $\frac{1}{3}$ while $\Delta_m=0$.

\begin{figure}[ht]
\centering
\includegraphics[width=7cm]{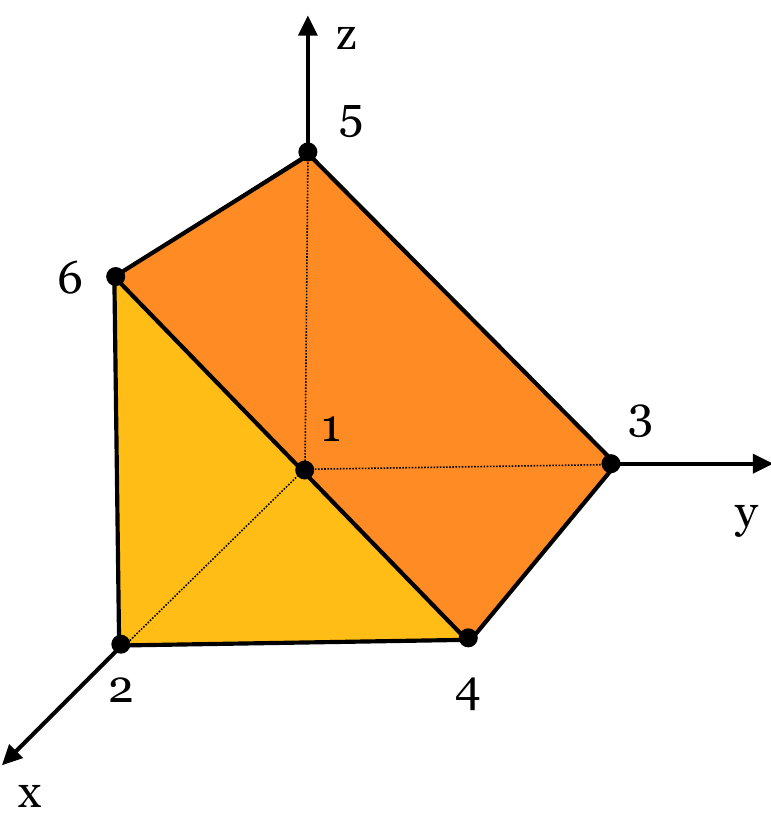}
\caption{Three dimensional toric diagram for D$_3$.}
\label{D3toric}
\end{figure}

We conclude by comparing the off-shell behavior of the partition function obtained from the application of the GK duality against the  expectations from the geometry. The equivalence of the two extremization problems has been already obtained in \cite{Jafferis:2011zi}, here we are interested in such matching from the perspective of the chiral quiver with four rank $N$ gauge nodes.
We start from the toric diagram given in Figure \ref{D3toric}.
The 4d vectors that specify toric diagram are
\begin{equation}
\label{toricD3}
G_{t}^{D_3}
=
\left(
\begin{array}{cccccc}
 0 & 1 & 0 & 1 & 0 & 1 \\
 0 & 0 & 1 & 1 & 0 & 0 \\
 0 & 0 & 0 & 0 & 1 & 1 \\
  1 & 1 & 1 & 1 & 1 & 1 \\
  \hline
 X_{21} & X_{12} & X_{14} & X_{23} & X_{42} & X_{31} \\
\end{array}
\right)
\end{equation}
where in the last line we specify the corresponding PM.
The volumes are expressed in terms of the Reeb vector as
\begin{eqnarray}\label{oparD3}
&&
\text{Vol}(\Sigma_1) =\frac{1}{b_1 b_2 b_3}, \quad 
\text{Vol}(\Sigma_2) =-\frac{1}{(b_1-4) b_2 b_3}\nonumber \\
&&
\text{Vol}(\Sigma_3) =-\frac{1}{b_1 b_3 (b_2+b_3-4)}
,\quad
\text{Vol}(\Sigma_4) =\frac{1}{(b_1-4) b_3 (b_2+b_3-4)} \nonumber \\
&& 
\text{Vol}(\Sigma_5) =-\frac{1}{b_1 b_2 \left(b_2+b_3-4\right)},\quad
\text{Vol}(\Sigma_6) =\frac{1}{\left(b_1-4\right) b_2 \left(b_2+b_3-4\right)}
\end{eqnarray}
such that 
\begin{eqnarray}
Z_{\text{MSY}} =
\frac{16}{\left(b_1-4\right) b_1 b_2 b_3 \left(b_2+b_3-4\right)}
\end{eqnarray}
The geometric R-charges are
 $\Delta_{i} = \frac{2 \text{Vol}(\Sigma_i)}{Z_{\text{MSY}} }$ and the Volume is 
 $\text{Vol}=\frac{\pi^4}{12} Z_{\text{MSY}} $.
 From the parameterization \eqref{oparD3} one then observes that 
\begin{equation}
\label{D3vol}
 \text{Vol}_{D_3} = \frac{\pi^4}{24}  \frac{1}{(\Delta_1+\Delta_2)(\Delta_3+\Delta_4)(\Delta_5+\Delta_6)(\Delta_1+\Delta_3+\Delta_5)(\Delta_2+\Delta_4+\Delta_6)}
\end{equation}
This off-shell function reproduces the volume once the charges are parameterized in terms of the Reeb vector.
Furthermore it is exactly the off-shell function expected from the free energy (\ref{D3nostro})  once the charges of the PM are specified in terms of the charges of the fields and of the monopoles and the constraints obtained by removing the 
flat directions are imposed. 
%
%
%
%
\subsection{The free energy of $\mathcal{C} \times \mathbb{C}$ and GK duality}
\label{subsec:CtimesC}
%
%
%
%

\begin{figure}[H] 
\begin{center}
\includegraphics[width=8cm]{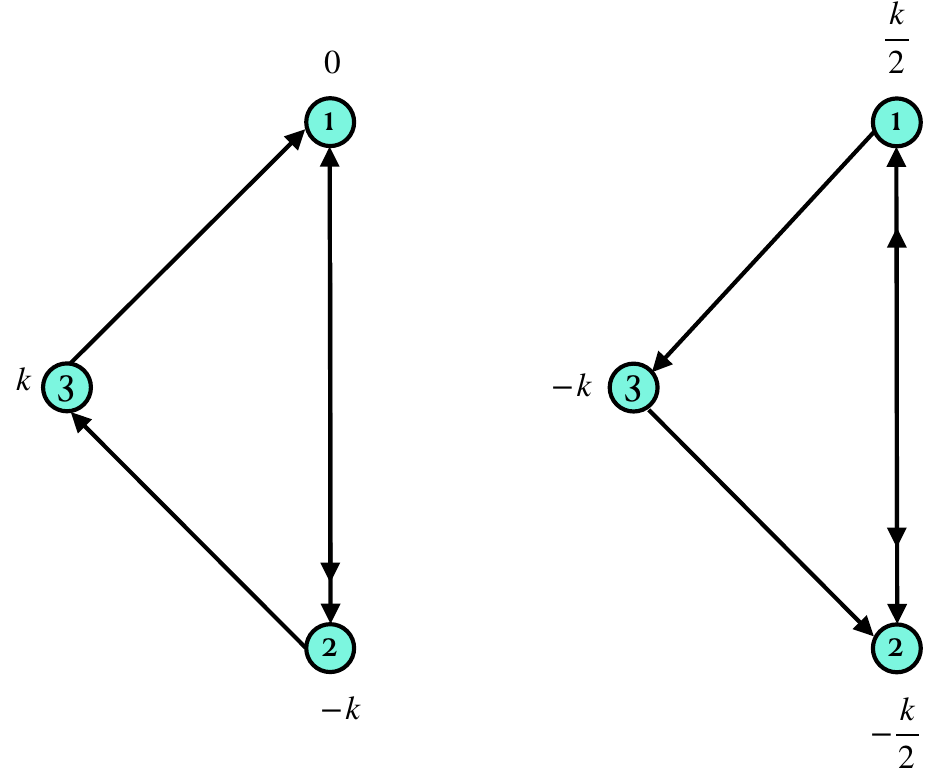}
\caption{Chiral quiver for the $\mathcal{C} \times \mathbb{C}$ geometry and its dual after the application of \eqref{GKrule}.}
\label{CtimesConifold}
\end{center} \end{figure}

The second example considered in this section corresponds to a model that has not been  studied  in \cite{Hosseini:2025jxb}, implying that our results are new and are a prediction that can be confirmed at numerical level.
We start considering the chiral quiver on the LHS of Figure \ref{CtimesConifold}.
The model in this case has superpotential
\begin{equation}
\label{WeleCC}
W = X_{12}Y_{23}Y_{31}Y_{12}X_{21} - 
X_{12} Y_{12} Y_{23}Y_{31}X_{21}
\end{equation}
where we consider an $U(N)$ gauge group for the node $3$ and a $SU(N)$ 
 gauge group for the other two nodes (in addition to the gauging of the diagonal $U(1)$).
The choice of the CS levels is spelled out in Figure \ref{CtimesConifold} as well.
We then dualize the node $3$, obtaining the quiver on the right of Figure \ref{CtimesConifold} with superpotential
\begin{equation}
\label{WeleCCdual}
W = X_{12}Y_{21}Y_{12}X_{21} - 
X_{12} Y_{12} Y_{21} X_{21}+Y_{21} Y_{13}Y_{32}
\end{equation}
Here we do not right down the explicit formulas for the partition function, but we only comment on the results that follows from the application of \eqref{GKrule}.
The new CS levels are given again in  Figure \ref{CtimesConifold}, while the  
$U(N)$ gauge group in the third node of the quiver becomes $U(|k|)$.
The duality dictionary can be read from the partition function as well, we have
$\Delta_{Y_{21}}=\Delta_{Y_{23}}+\Delta_{Y_{31}} $, $\Delta_{Y_{13}} = 1-\Delta_{Y_{31}}$ and $\Delta_{Y_{32}} = 1-\Delta_{Y_{23}}$.
Furthermore the monopole charge $\tilde \Delta_m  = \tilde \Delta_{m_1} +\tilde \Delta_{m_2} $  is shifted by the charges of the fields $\Delta_{Y_{23}}$ and $\Delta_{Y_{31}}$ of the quiver with three rank-N gauge nodes. This shift is crucial in the analysis of the free energy and it is 
\begin{equation}
\label{momo}
2\tilde \Delta_{m_1}=2\Delta_{m_1} +\Delta_{m_3} + k \Delta_{Y_{31}}, \quad
2\tilde \Delta_{m_2}=2\Delta_{m_2} +\Delta_{m_3} - k \Delta_{Y_{23}}
\end{equation}
Here we restrict to the case $k=1$, where  the leading $N^{3/2}$ contribution to the free energy has been obtained in \cite{Jafferis:2011zi} and it is
\begin{equation}
\label{t11ckleb}
F = \frac{2\sqrt{2}\pi N^{\frac{3}{2}} }{3} 
\sqrt{\frac{\hat\Delta_{Y_{21}}
(\hat\Delta_{X_{21}}+\hat\Delta_{Y_{12}})(\hat\Delta_{X_{12}}+\hat\Delta_{X_{21}})
(\hat\Delta_{Y_{21}}+2\hat\Delta_{Y_{12}})(\hat\Delta_{Y_{21}}+2\hat\Delta_{X_{12}})
}
{4-
\hat\Delta_{Y_{21}}
}}
\end{equation}
where $\hat\Delta_{Z_{21}} =\Delta_{Z_{21}}-2 \tilde \Delta_m $ and $\hat\Delta_{Z_{12}} =\Delta_{Z_{12}}+2 \tilde \Delta_m $, with $Z = \{X,Y\}$.
We than use the duality dictionary to obtain the free energy of the chiral version of $\mathcal{C} \times \mathbb{C}$ with three gauge groups
of rank $N$. This is obtained by substituting in (\ref{t11ckleb}) the relation
 $\Delta_{Y_{21}}=\Delta_{Y_{23}}+\Delta_{Y_{31}}$ 
in addition to (\ref{momo}). We obtain
\begin{eqnarray}
\label{t11cnoi}
F = \frac{4\sqrt{2}\pi N^{\frac{3}{2}} }{3} &&
\sqrt{\frac{1}{\Delta _m+2-\Delta _{Y_{23}}}
(\Delta _{X_{12}}+\Delta _{X_{21}}) 
(\Delta _{Y_{23}}- \Delta _m) (\Delta _m+\Delta _{Y_{12}}+\Delta _{Y_{31}}) }\nonumber \\ \times &&
\sqrt{
(\Delta _{X_{21}}+\Delta _{Y_{12}}) 
(\Delta _m+\Delta _{X_{12}}+\Delta _{Y_{31}})}
\end{eqnarray}
where $\Delta_m = \sum_{a=1}^3 \Delta_{m_a}$.

\begin{figure}[H] 
\begin{center}
\includegraphics[width=8cm]{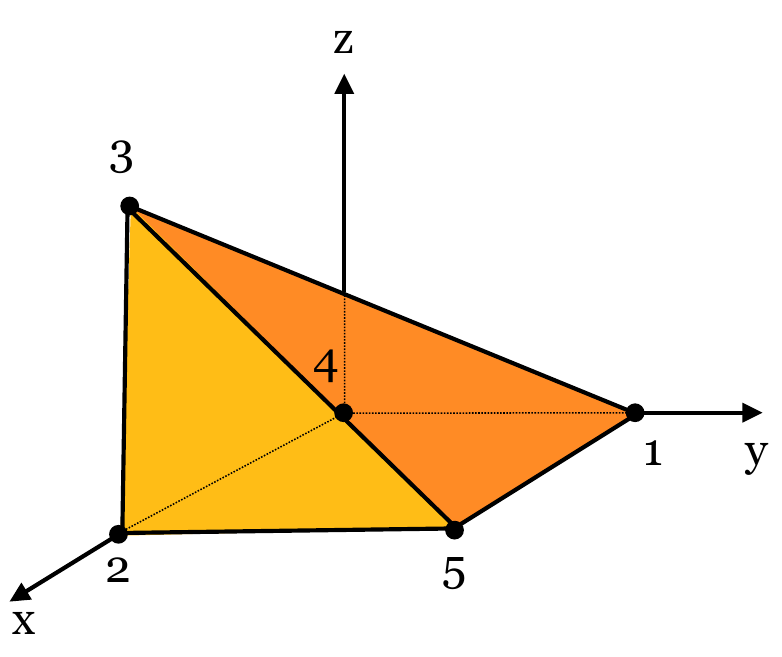}
\caption{Toric diagram for the $\mathcal{C} \times \mathbb{C}$ geometry.}
\label{Fig:CtimesCToric}
\end{center} \end{figure}

We conclude by comparing the off-shell behavior of the partition function obtained from the application of the GK duality against the  expectations from the geometry. The equivalence of the two extremization problems has been  obtained in \cite{Jafferis:2011zi}, here we are interested in such matching from the perspective of the chiral quiver with three rank-$N$ gauge nodes.
The $Z_{MSY}$ can be computed starting from the toric diagram in Figure
\ref{Fig:CtimesCToric}. 
The volumes are expressed in terms of the Reeb vector as
\begin{eqnarray}\label{opar}
\text{Vol}(\Sigma_1) &=& -\frac{1}{\left(b_1-b_3\right) b_3 \left(b_2+b_3-4\right)}  \nonumber \\
\text{Vol}(\Sigma_2) &=&-\frac{1}{\left(b_1-4\right) b_2 b_3} \nonumber \\
\text{Vol}(\Sigma_3) &=&\frac{4-b_3}{\left(b_1-4\right) b_2 \left(b_1-b_3\right) \left(b_2+b_3-4\right)} \nonumber \\
\text{Vol}(\Sigma_4) &=& \frac{1}{b_1 b_2 b_3-b_2 b_3^2}\nonumber \\
\text{Vol}(\Sigma_5) &=&\frac{1}{\left(b_1-4\right) b_3 \left(b_2+b_3-4\right)}
\end{eqnarray}
such that 
\begin{eqnarray}
Z_{\text{MSY}} =
-\frac{4 \left(b_3-4\right)}{\left(b_1-4\right) b_2 \left(b_1-b_3\right) b_3 \left(b_2+b_3-4\right)}
\end{eqnarray}
The geometric R-charges are
 $\Delta_{i} = \frac{2 \text{Vol}(\Sigma_i)}{Z_{\text{MSY}} }$ and the Volume is 
 $\text{Vol}=\frac{\pi^4}{12} Z_{\text{MSY}} $.
 The Volume  function can be equivalently expressed in terms of the charges $\Delta_i$ as
 \begin{eqnarray}
 \label{T11Cvol}
\text{Vol}_{\mathcal{C} \times \mathbb{C}} = \frac{ \pi^4 }{48}
 \frac{2-\Delta _3}{\Delta _3 \left(\Delta _1+\Delta _4\right) \left(\Delta _2+\Delta _4\right) \left(\Delta _1+\Delta _5\right) \left(\Delta _2+\Delta _5\right)}
 \end{eqnarray}
 This off-shell function reproduces the volume once the charges are parameterized in terms of the Reeb vector.
Furthermore it is exactly the off-shell function expected from the free energy (\ref{t11cnoi})  once the charges of the PM are specified in terms of the charges of the fields and of the monopoles and the constraints obtained by removing the 
flat directions are imposed. 
 In this case the  PM $\Pi_{i}$ associated to the lattice points of the toric diagram when considering the first quiver in Figure \ref{CtimesConifold} are
\begin{equation}
\vec{\Pi} = \{X_{21},Y_{31},Y_{23},Y_{12},X_{12}\} 
\end{equation}

%
%
%
%
\subsection{The free energy of $Q^{111}$ and GK duality}
\label{subsec:Q111}
%
%
%
%

The next example under investigation is the chiral quiver associated to  $Q^{111}$ .
This is  the  first case recently studied in 
\cite{Hosseini:2025jxb} by a numeric evaluation of the large $N$ three sphere partition function.

It follows that we consider the GK duality for the nodes with $N_f=N_c$ of the $Q^{111}$ model introduced above. We focus on the phase studied in \cite{Hosseini:2025jxb}, with CS levels 
$(k_1,k_2,k_A,k_B) = (0,0,1,-1)$, even if other CS assignments can give rise to the same moduli space (see e.g. \cite{Franco:2009sp,Amariti:2009rb,Aganagic:2009zk}). For the ease of notation in the following we rename the nodes $N_{A,B}$ as $N_{3,4}$. The three sphere partition function corresponds to the following matrix integral

\begin{eqnarray}
\label{chiralQ111}
Z_{Q^{111}}\!&\!=\!&\!\frac{1}{(N!)^4}
\int 
\frac{ \prod_{i=1}^{N} \prod_{a=1}^4
d \sigma_{i,a}
e^{i  \pi \sum_{i=1}^N ( \sigma_{i,3}^2-\sigma_{i,4}^2)+2  \pi \sum_{a=1}^{4} \sum_{i=1}^N \Delta_{m_a} \sigma_{i,a}}}{
\prod_{a=1}^4 \prod_{1\leq i < j \leq N} \left(2 \sinh^{-2}(\pi (\sigma_{i,a}-\sigma_{j,a}))\right)^{-2}} \nonumber \\
&&
\prod_{i,j=1}^{N} 
e^{\sum_{k=1}^2\ell(1- \Delta_{12}^{(k)}+i (\sigma_{i,1} - \sigma_{j,2}))+\ell(1-\Delta_{23}+i(\sigma_{i,2} -\sigma_{j,3} ))+\ell(1-  \Delta_{24} + i(\sigma_{i,2} -\sigma_{j,4}))}
\nonumber \\
&&
e^{\ell(1- \Delta_{31}+i(\sigma_{i,3} -\sigma_{j,1} )) + \ell(1- \Delta_{41}+i(\sigma_{i,4} -\sigma_{j,1})) }
\delta\Big( \sum_{i=1}^{N} (\sigma_{i,1}-\sigma_{i,2})\Big) 
\end{eqnarray}

This partition function is almost identical to the one recently studied at large $N$ in \cite{Hosseini:2025jxb}, where the scaling $N^{3/2}$ has been obtained after more that a decade.
The main difference between the two models stays in the way we treated the $U(1)$ factors here, indeed in \cite{Hosseini:2025jxb} all the FI terms have been set to zero explicitely in the matrix model, while here we are keeping the FI of $U(N)_3$ and $U(N)_4$ in addition to the one of the diagonal $U(1) \subset U(N_1) \times U(N_2)$. We eliminated the one of the anti-diagonal $U(1) \subset U(N_1) \times U(N_2)$, by integrating over it, or equivalently by adding the delta function.
These differences do not alter the large N scaling but they are relevant in order to claim the existence of a Large $N$ duality beyond the matching of the three sphere free energies.

Next we provide the main argument in favour of the $N^{3/2}$ scaling that does not require any new evaluation but that, thanks to the application of \eqref{GKrule}
allows us to rewrite the partition function \eqref{chiralQ111} as 

\begin{eqnarray}
\label{dualQ111}
Z_{Q^{111}}\!&\!=\!&\!
\int
\frac
{ 
\prod_{i=1}^{N} \prod_{a=1}^2 d \sigma_{i,a} d\sigma_{3} d\sigma_{4}
e^{-i  \pi ( \sigma_{i,3}^2-\sigma_{i,4}^2)-2 \pi \left( \Delta_{m_3} \sigma_{3}+ \Delta_{m_4}  \sigma_{4}\right)+2  \pi  \sum_{i=1}^N   (  \tilde \Delta_{m_1} \sigma_{i,1}+  \tilde \Delta_{m_2} \sigma_{i,2} )}}
{\prod_{a=1}^2 \prod_{1\leq i < j \leq N} \left(2 \sinh(\pi (\sigma_{i,a}-\sigma_{j,a}))\right)^{-2}} \nonumber \\
&&
\prod_{i,j=1}^{N} 
e^{\sum_{k=1}^2 \ell (1-\Delta_{12}^{(k)} + i (\sigma_{i,1} - \sigma_{j,2} ) ) + \ell(1- \Delta_{24}-\Delta_{41} +i(\sigma_{i,2} - \sigma_{j,1})) }
\nonumber \\ &&
e^{\ell(1-\Delta_{23}-\Delta_{31}+i(\sigma_{i,2} - \sigma_{j,1}))
+
\ell(\Delta_{23} +i (\sigma_3 -\sigma_{i,2} ))
+
\ell(\Delta_{24} + i(\sigma_{4} -\sigma_{i,2})))
}
\nonumber \\ &&
e^{\ell(\Delta_{31}+i(\sigma_{j,1}-\sigma_{3})+\ell( \Delta_{41}+i(\sigma_{j,1} -\sigma_{4} ))}
\delta\big(\sum_{i=1}^{N} (\sigma_{i,1}-\sigma_{i,2})\big) 
\end{eqnarray}
where $\tilde \Delta_{m_{1,2}} $ are shifted following  \eqref{GKrule} and they become
\begin{equation}
\label{shiftedFI}
\begin{array}{l}
		 2\tilde\Delta_{m_1}=\Delta_{41}-\Delta_{31}+
		 \Delta_{m_3}+\Delta_{m_4}+2\Delta_{m_1}
		 \\
			 2\tilde\Delta_{m_2}=\Delta_{23}-\Delta_{24}+\Delta_{m_3}+\Delta_{m_4}+2\Delta_{m_2} 
			 \end{array}
		 \end{equation}
Furthermore the symmetries of the quiver yield  to 
\begin{equation}
\Delta_{12}^{(1)} = \Delta_{23} =  \Delta_{31} \equiv \Delta_1,
 \quad 
\Delta_{12}^{(2)} = \Delta_{24} =  \Delta_{41} \equiv \Delta_2
\end{equation}
with $\Delta_1+\Delta_2 = \frac{2}{3}$.

\begin{figure}
\begin{center}
  \includegraphics[width=7cm]{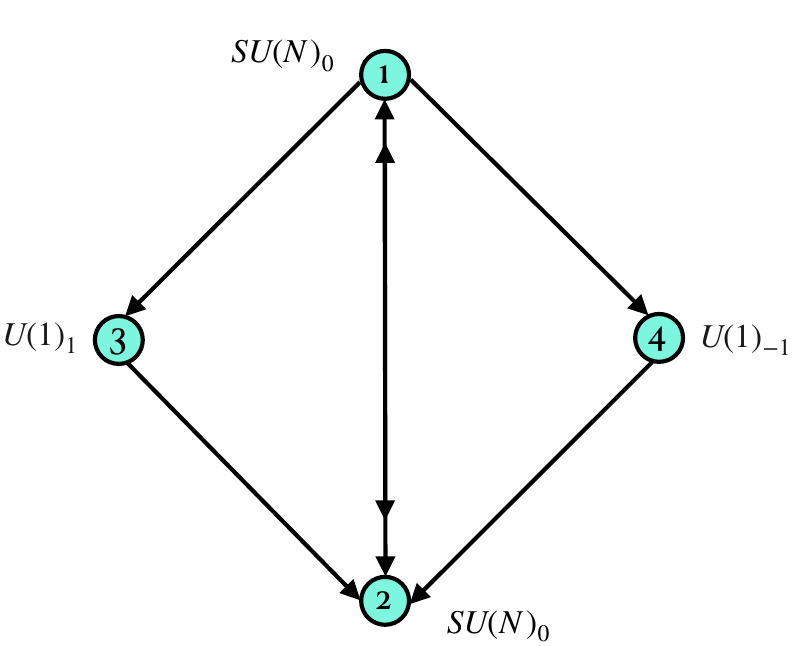}
  \end{center}
  \caption{Quiver obtained after Giveon-Kutasov duality 
  	on the chiral quiver describing the $Q^{1,1,1}$ model. We did not represent the extra $U(1)_{diag.}$ gauging in the picture.
    }
    \label{Fig:GKdualQ111}
\end{figure}

At this point we can study the large $N$ matrix model for their dual phase just obtained at the level of exact mathematical identities. This corresponds to an $U(1)^3 \times SU(N)^2$ quiver, where the two original $U(N)_{k=\pm 1}$ factors associated to the nodes $3$ and $4$ in the quiver are now $U(1)_{k=\mp 1}$.
The matrix model is compatible with the superpotential 
\begin{equation}
W = X_{12}^{(1)} X_{21}^{(1)} 
X_{12}^{(2)} X_{21}^{(2)} 
-
X_{12}^{(1)} X_{21}^{(2)} 
X_{12}^{(2)} X_{21}^{(1)} 
+
X_{21}^{(1)} X_{13} X_{32}
+
X_{21}^{(2)} X_{14} X_{42}
\end{equation}
associated to the quiver in Figure \ref{Fig:GKdualQ111}, where
$X_{21}^{(1)} = X_{23} X_{31} $ and $X_{21}^{(2)} = X_{24} X_{41} $.
At the level of the $R$ symmetry the dual charges of the bifundamentals 
$ \Delta_{23} +  \Delta_{31} =  \Delta_{21}^{(1)}  \equiv 2 \Delta_1 $
and
$ \Delta_{24} +  \Delta_{41} =  \Delta_{21}^{(2)}  \equiv 2 \Delta_2$. This gives also
$ \Delta_{13} + \Delta_{32} = 2 -2 \Delta_1$ and $ \Delta_{14} + \Delta_{42} = 2 -2 \Delta_2$.

 In this case the free energy has been computed in \cite{Jafferis:2011zi} with the simplifications
 $\Delta_{X_{12}^{(1)}} =\Delta_{X_{12}^{(2)}} = 1-\Delta$ and
  $\Delta_{X_{21}^{(1)}} =\Delta_{X_{21}^{(2)}} =\Delta$, which are valid at the extremum.
  The free energy is only a function of the topological charge $\Delta_T - \Delta_{\tilde T} =2\Delta_m$.
  In this case the effective FI terms is given in \eqref{shiftedFI} in terms of the charges of the original fields.
  The free energy is then \cite{Jafferis:2011zi}
\begin{equation}
F_{S^3} = \frac{4 \pi N^{3/2}}{3} \frac{|1-\tilde \Delta_m^2  |}{\sqrt{3-\tilde\Delta_m^2 }}
\end{equation}  
and there is a further constraint from the flat direction given by $2/(3-\tilde\Delta_m^2)=\Delta$.
At the fixed point the charge $\tilde \Delta_m$ that maximizes the free energy is $\Delta_m=0$ while the charge $\Delta$ is constrained by the symmetries and by the flat directions and at the fixed point is  $\Delta  =2/3$.
  
Using the symmetries and the duality dictionary of the duality we have
\begin{equation}
\Delta_{24}-\Delta_{41}=\Delta_{23} - \Delta_{31} =0 ,\quad 
\Delta_{24} + \Delta_{41} =\Delta_{23} + \Delta_{31} = \Delta
 \end{equation}
It follows that the charges of the six fields for the chiral quiver are all equal to $\frac{1}{3}$ at the fixed point.

In this case we can have a deeper understading of the off-shell relation between the chiral quiver with four gauge nodes and the quiver with chiral flavor by studying the relation with the volume computation.
 The toric diagram has been given in formula \eqref{torcDQ111}. 
The volumes expressed in terms of the components of the Reeb vector are
\begin{eqnarray}
\text{Vol}(\Sigma_1)=-\frac{b_1+b_3}{b_1 \left(b_2-4\right) b_3 \left(b_1+b_2+b_3-4\right)},
 \nonumber \\
\text{Vol}(\Sigma_2)=-\frac{b_1+b_2}{b_1 b_2 \left(b_3-4\right) \left(b_1+b_2+b_3-4\right)},
 \nonumber \\
\text{Vol}(\Sigma_3)=\frac{b_1+b_3-8}{\left(b_1-4\right) b_2 \left(b_3-4\right) \left(b_1+b_2+b_3-8\right)},
 \nonumber \\
\text{Vol}(\Sigma_4)=\frac{b_1+b_2-8}{\left(b_1-4\right) \left(b_2-4\right) b_3 \left(b_1+b_2+b_3-8\right)},
 \nonumber \\
\text{Vol}(\Sigma_5)=\frac{b_2+b_3-8}{b_1 \left(b_2-4\right) \left(b_3-4\right) \left(b_1+b_2+b_3-8\right)},
 \nonumber \\
\text{Vol}(\Sigma_6)=-\frac{b_2+b_3}{\left(b_1-4\right) b_2 b_3 \left(b_1+b_2+b_3-4\right)}\,
\end{eqnarray}
Summing over these volumes we obtain function $Z_{MSY}$, which si given by
\begin{equation}
Z_{MSY} = \frac{8 \left(\left(b_2 \left(b_3-2\right)-2 b_3\right) b_1^2+\left(b_2+b_3-8\right) \left(b_2 \left(b_3-2\right)-2 b_3\right) b_1-2 b_2 b_3 \left(b_2+b_3-8\right)\right)}{\left(b_1-4\right) b_1 \left(b_2-4\right) b_2 \left(b_3-4\right) b_3 \left(b_1+b_2+b_3-8\right) \left(b_1+b_2+b_3-4\right)}
\end{equation}
Various expressions reproducing this function have been proposed in the literature based on the charges 
$\Delta_{\pi_i} = \frac{2\text{Vol}(\Sigma_i)}{Z_{MSY}}$, which are 
constrained by $\sum_i \Delta_{\pi_i} =2 $.
Observe that these variables satisfy another pair of non linear relations
\begin{equation}
\label{q111c1}
\Delta _{\pi _6}+\frac{\left(\Delta _{\pi _1} \Delta _{\pi _3}-\Delta _{\pi _2} \Delta _{\pi _4}\right) \left(\Delta _{\pi _1}+\Delta _{\pi _2}+\Delta _{\pi _3}+\Delta _{\pi _4}+2 \Delta _{\pi _5}\right)}{\left(\Delta _{\pi _3}-\Delta _{\pi _4}\right) \Delta _{\pi _5}+\Delta _{\pi _1} \left(2 \Delta _{\pi _3}+\Delta _{\pi _5}\right)-\Delta _{\pi _2} \left(2 \Delta _{\pi _4}+\Delta _{\pi _5}\right)}=0
\end{equation}
and
\begin{equation}
\label{q111c2}
\Delta _{\pi _2}+\frac{\left(\Delta _{\pi _1}+\Delta _{\pi _3}+2 \Delta _{\pi _4}+\Delta _{\pi _5}+\Delta _{\pi _6}\right) \left(\Delta _{\pi _5} \Delta _{\pi _6}-\Delta _{\pi _1} \Delta _{\pi _3}\right)}{\Delta _{\pi _4} \left(\Delta _{\pi _5}-\Delta _{\pi _3}\right)+\left(\Delta _{\pi _4}+2 \Delta _{\pi _5}\right) \Delta _{\pi _6} -\Delta _{\pi _1} \left(2 \Delta _{\pi _3}+\Delta _{\pi _4}\right)}=0
\end{equation}
Such non linear relations can be obtained directly as discussed in \cite{Hosseini:2019ddy} from the quartic formula of \cite{Amariti:2011uw,Amariti:2012tj}.

The parameterization that reproduces the extremization problem discussed on the field theory side is
\begin{equation}
\begin{array}{ccc}
\Delta _{\pi _6}= 1-\Delta&\quad \quad 
\Delta _{\pi _1}  = \frac{\Delta }{2}-\frac{\Delta _m}{2}& \quad \quad
\Delta _{\pi _2}= \frac{\Delta }{2}+\frac{\Delta _m}{2}\\
\Delta _{\pi _4}= \frac{\Delta }{2}-\frac{\Delta _m}{2}&\quad\quad 
\Delta _{\pi _5}= 1-\Delta &\quad\quad
\Delta _{\pi _3}= \frac{\Delta }{2}+\frac{\Delta _m}{2}
\end{array}
\end{equation}
Observe that the  constraint \eqref{q111c1} is satisfied automatically by this parameterization while the constraint \eqref{q111c2}
 is satisfied if  $\Delta  (\Delta _m^2-3)+2=0$.
 
 In order to connect with the results of \cite{Hosseini:2025jxb} here we discuss a different parameterization (see \cite{Kim:2012vza})
\begin{equation}
\label{parcoreans}
\begin{array}{ccc}
\Delta _{\pi _6}= \Delta _1&\quad \quad 
\Delta _{\pi _1}= \Delta _2-\frac{\Delta _m}{2}&\quad \quad 
\Delta _{\pi _2}= \Delta _2+\frac{\Delta _m}{2}\\
\Delta _{\pi _4}= \Delta _1-\frac{\Delta _m}{2}&\quad \quad 
\Delta _{\pi _5}= \Delta _2&\quad \quad 
\Delta _{\pi _3}= \Delta _1+\frac{\Delta _m}{2}
\end{array}
\end{equation}
Using this parametrization the free energy is 
\begin{equation}
Z_{MSY} = 
\frac{\left(2 \Delta _1+\Delta _2\right){}^2 \left(\Delta _1+2 \Delta _2\right){}^2-\left(\Delta _1^2+\Delta _2 \Delta _1+\Delta _2^2\right) \Delta _m^2}{6 \Delta _1 \Delta _2 \left(2 \Delta _1+\Delta _2\right) \left(\Delta _1+2 \Delta _2\right) \left(\left(2 \Delta _1+\Delta _2\right){}^2-\Delta _m^2\right) \left(\left(\Delta _1+2 \Delta _2\right){}^2-\Delta _m^2\right)}
\end{equation}
The two constraints spelled out above are satisfied if $\Delta _1=\Delta _2$ and $\Delta _m=0$ in addition to the constraint $3 \left(\Delta _1+\Delta _2\right)=2$. 

We can see that the duality dictionary discussed above is consistent with the volume computations. On the quivwer with chiral flavor we can indeed  identify the charges  
as \cite{Benini:2009qs}
$\Delta_{\pi_6} =\Delta_{X_{12}}^{(1)}$,
$\Delta_{\pi_1}+\Delta_{\pi_4} =     \Delta_{\tilde T}$,
$\Delta_{\pi_3} +\Delta_{\pi_4} =     \Delta_{X_{21}^{(2)}}$,
$\Delta_{\pi_2} +\Delta_{\pi_3}=     \Delta_{T}$,
$\Delta_{\pi_5} =\Delta_{X_{12}}^{(2)}$ and
$\Delta_{\pi_1}+\Delta_{\pi_2} =     \Delta_{X_{21}^{(2)}}$.
while consistently  the charge assignation for the quiver with four gauge nodes is
$\Delta_{\pi_6} =\Delta_{X_{12}}^{(1)}$,
$\Delta_{\pi_4} =      \Delta_{X_{24}}-\frac{\Delta _m}{2}$,
$\Delta_{\pi_3} =     \Delta_{X_{41}}+\frac{\Delta _m}{2}$,
$\Delta_{\pi_2} =     \Delta_{X_{23}}-\frac{\Delta _m}{2}$,
$\Delta_{\pi_5} =\Delta_{X_{12}}^{(2)}$ and
$\Delta_{\pi_1} =     \Delta_{X_{31}}+\frac{\Delta _m}{2}$.

%
%
%
%
\subsection{The free energy of the cubic conifold and GK duality}
\label{cubiccon}
\begin{figure}[H] 
\begin{center}
\includegraphics[width=6cm]{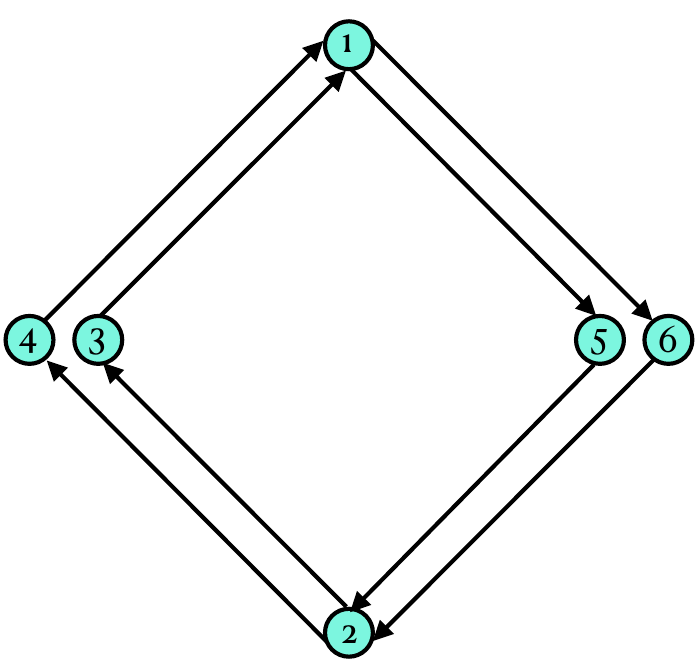}
\caption{Chiral quiver associated to the cubic conifold. The CS levels in this case are $\vec k = \{0,0,1,-1,1,-1\}$ }
\label{Fig:cubicele}
\end{center} \end{figure}

%
%
%
%
%
The last example studied in this section corresponds to the "cubic conifold".
The model was originally studied in \cite{Benini:2011mf} by flavoring the ABJM theory. In absence of such flavoring other construction of the same geometry are possible, both with chiral and non-chiral quivers.
Here we focus on the chiral quiver reported on Figure \ref{Fig:cubicele}.
The gauge group is $SU(N_1)_{k_1} \times SU(N_2)_{k_2}  \times U(1) \times  \prod_{i=3}^6 U (N_i)_{k_i}$ with $N_i=N$ and  $\vec k = \{0,0,1,-1,1,-1\}$.
The abelian factor corresponds to the diagonal $U(1) \subset U(N_1) \times U(N_2)$. 
The superpotential for this phase is
\begin{equation}
W = X_{23} X_{31}X_{15}X_{52} X_{24} X_{41} X_{16} X_{62} - X_{23} X_{31}X_{16}X_{62} X_{24} X_{41} X_{15} X_{52}
\end{equation}

\begin{figure}[H] 
\begin{center}
\includegraphics[width=6cm]{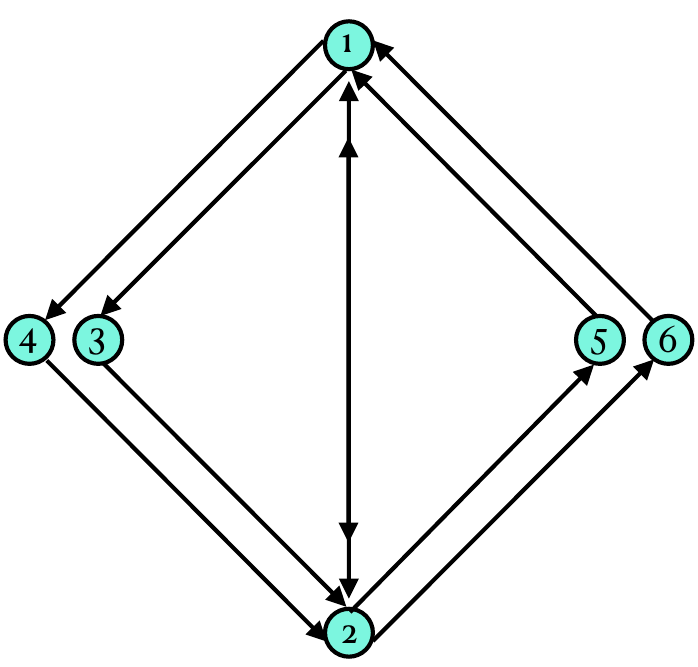}
\caption{Dual chiral quiver associated to the cubic conifold. In this case the gauge groups $3,4,5$ and $6$ are abelian.}
\label{Fig:cubicmag}
\end{center} \end{figure}

The dual theory is obtained by applying on the three sphere partition function 
the relation \eqref{GKrule} sequentially on $U(N_{3,4,5,6})$.
The four dual gauge groups correspond either to $U(1)_{1}$ or to $U(1)_{-1}$, while the CS levels of the groups $SU(N_1)$ and $SU(N_2)$ (and of the diagonal $U(1)$) are still vanishing.
The dual quiver is represented in Figure  \ref{Fig:cubicmag}.
The superpotential for this dual phase is 
\begin{eqnarray}
\label{cubicdual}
W &=& Y_{13} Y_{32} {X_{21}}^{(1)} +Y_{14} Y_{42} X_{21}^{(2)} +
Y_{25} Y_{51} {X_{12}}^{(1)} +Y_{26} Y_{61} {X_{12}}^{(2)} \nonumber \\
&+&
 {X_{12}}^{(1)} {X_{21}}^{(1)}{X_{12}}^{(2)} {X_{21}}^{(2)}-{X_{12}}^{(1)} {X_{21}}^{(2)}{X_{12}}^{(1)} {X_{21}}^{(2)}
\end{eqnarray}
The leading $N^{3/2}$ contribution to the free energy has been obtained in \cite{Jafferis:2011zi} and it is
\begin{equation}
F =\frac{2 \pi N^{3/2}}{3} \sqrt{
\prod_{j,k=1,2} 2(\Delta_{X_{12}^{(i)}} + \Delta_{X_{21}^{(j)}}  ) (1-{\tilde \Delta_m}^2)
}
\end{equation}
with $\tilde \Delta_m = \tilde \Delta_{m_1} +\tilde \Delta_{m_2}$.
We can extract the free energy for the chiral theory with six rank-N gauge groups
by exploiting the duality dictionary implied by \eqref{GKrule}.
 Explicitly there are four relations associated to the bifundamentals.
They are
$\Delta_{X_{12}^{(1)}} = \Delta_{X_{13}} + \Delta_{X_{32}}$, 
$\Delta_{X_{12}^{(2)}} = \Delta_{X_{14}} + \Delta_{X_{42}}$,
$\Delta_{X_{21}^{(1)}} = \Delta_{X_{25}} + \Delta_{X_{51}} $ and 
$\Delta_{X_{21}^{(1)}} = \Delta_{X_{26}} + \Delta_{X_{61}} $.
In addition the effective FI terms of the dual theory, $\tilde \Delta_{m_{1,2}}$, are obtained from (\ref{betterFI})  
and it boils down to the relation 
$2 \tilde \Delta_m =2 \Delta_m +  (\Delta_{X_{13}} - \Delta_{X_{32}} )
+ ( \Delta_{X_{14}} - \Delta_{X_{42}} )
+ (\Delta_{X_{25}} - \Delta_{X_{51}})+
(\Delta_{X_{26}} - \Delta_{X_{61}})$, where 
$\Delta_m = \sum_{a=1}^{6} \Delta_{m_a}$.
We then obtain the free energy in the electric theory by using these relations.
We have
\begin{eqnarray}
\label{elecubicconifoldfree}
F &=&\frac{4 \pi N^{3/2}}{3}  \sqrt{
 \left(\Delta _{X_{13}}+\Delta _{X_{25}}+\Delta _{X_{32}}+\Delta _{X_{51}}\right) 
\left(\Delta _{X_{14}}+\Delta _{X_{25}}+\Delta _{X_{42}}+\Delta _{X_{51}}\right)} \nonumber \\
&\times &
\sqrt{\left(\Delta _{X_{13}}+\Delta _{X_{26}}+\Delta _{X_{32}}+\Delta _{X_{61}}\right) 
\left(\Delta _{X_{14}}+\Delta _{X_{26}}+\Delta _{X_{42}}+\Delta _{X_{61}}\right) }  \\
&\times &
\sqrt{
(\Delta _{X_{13}}\!+\!\Delta _{X_{14}}\!+\!\Delta _{X_{25}}\!+\!\Delta _{X_{26}} \!+\! \Delta_{m}) 
(\Delta _{X_{32}}\!+\!\Delta _{X_{42}}\!+\!\Delta _{X_{51}}\!+\!\Delta _{X_{61}}\!-\! \Delta_{m}) 
}\nonumber
\end{eqnarray}

\begin{figure}[H] 
\begin{center}
\includegraphics[width=8cm]{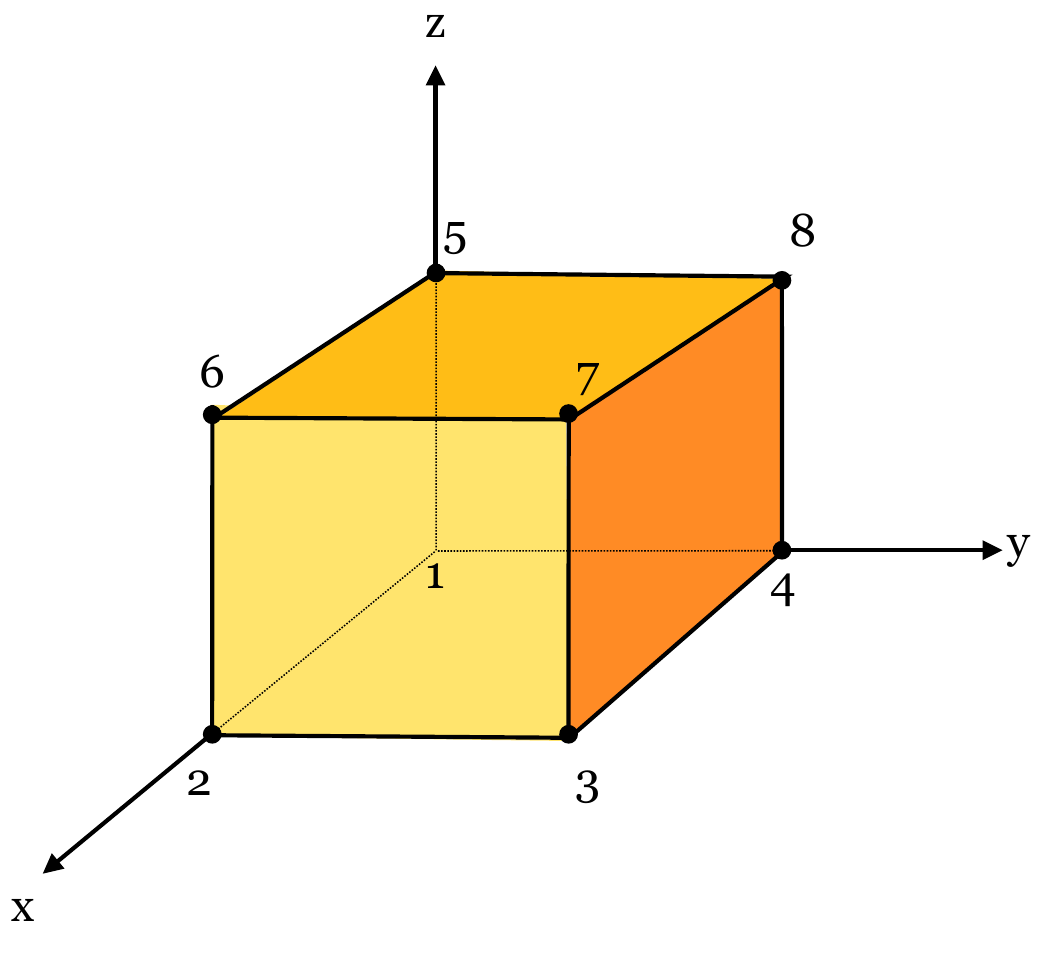}
\caption{Toric diagram for the cubic conifold}
\label{Fig:cubictoric}
\end{center} 
\end{figure}

We conclude the analysis by comparing this expression with the one expected from the from the geometry.
The toric diagram is given in Figure \ref{Fig:cubictoric}
and the volumes of the 5-cycles $\Sigma_i$ on which the M5 branes are wrapped are 
expressed in terms of the components of the Reeb vector as
\begin{equation}
\begin{array}{ll}
 \text{Vol}\left(\Sigma _1\right)=\frac{1}{b_1 b_2 b_3}\,, &\quad
 \text{Vol}\left(\Sigma _2\right)=\frac{1}{ (4-b_1)b_2 b_3}\,, \\
 \text{Vol}\left(\Sigma _3\right)=\frac{1}{(4-b_1) (4-b_2) b_3} \,,&\quad
 \text{Vol}\left(\Sigma _4\right)=\frac{1}{ b_1(4- b_2 )b_3}\, , \\
 \text{Vol}\left(\Sigma _5\right)=\frac{1}{b_1 b_2(4- b_3)} &\quad
 \text{Vol}\left(\Sigma _6\right)=\frac{1}{(4-b_1) b_2 (4-b_3)}\,, \\
 \text{Vol}\left(\Sigma _7\right)=\frac{1}{(4-b_1) (4-b_2) (4-b_3)}\,, &\quad
 \text{Vol}\left(\Sigma _8\right)=\frac{1}{b_1 (4-b_2) (4-b_3)} \, .
\end{array}
\end{equation}
and summing these volumes we obtain the $Z_{MSY}$ function.
Defining the  charges $\Delta_{\pi_i} =  \frac{2\text{Vol}\left(\Sigma _1\right)}{Z_{MSY}}$ we can reformulate the $Z_{MSY}$ function in terms of the charges 
$\Delta_{\pi_i}$. There are many equivalent expressions (indeed the charges satisfy various constraints). Here we focus on the following
\begin{equation}
\label{cubicconifoldcfocus}
 Z_{MSY} =
 \frac{1}{2 
\kappa _{1256} \kappa _{2367} \kappa _{1458} \kappa _{3478}   \kappa _{1234}  \kappa _{5678} }
 \end{equation}
where  $\kappa_{i j k \ell} \equiv \Delta _{\pi _i}+\Delta _{\pi _j}+\Delta _{\pi _k}+\Delta _{\pi _\ell}$.
 
 We can compare this last expression with the results obtained from the field theory analysis, focusing on the electric phase.
 In this case each PM in the geometry is associated to a bifundamental and the dictionary between the   charges of the 
fields $\Delta_{X_{ij}}$ and of the FI $\Delta_{m}$  and the charges $\Delta_{\pi_i}$ is 
\begin{equation}
\begin{array}{llll}
\Delta_{\pi_1} \!=\! \Delta_{X_{13}}\!-\!\frac{\Delta_{m}}{4 }, &\quad
\Delta_{\pi_2} \!=\!  \Delta_{X_{14}}\!-\!\frac{\Delta_{m}}{4 }, &\quad
\Delta_{\pi_3} \!=\!  \Delta_{X_{25}}\!-\!\frac{\Delta_{m}}{4 }, &\quad
\Delta_{\pi_4} \!=\!  \Delta_{X_{26}}\!-\!\frac{\Delta_{m}}{4 }, \\
\Delta_{\pi_5} \!=\!  \Delta_{X_{32}}\!+\!\frac{\Delta_{m}}{4 }, &\quad
\Delta_{\pi_6} \!=\!  \Delta_{X_{42}}\!+\!\frac{\Delta_{m}}{4 }, &\quad
\Delta_{\pi_7} \!=\!  \Delta_{X_{51}}\!+\!\frac{\Delta_{m}}{4 },  &\quad
\Delta_{\pi_8} \!=\!  \Delta_{X_{61}}\! +\!\frac{\Delta_{m}}{4 } \, .
\end{array}
\end{equation}
This parametrization makes the equivalence 
between the field energy read from the partition function (\ref{elecubicconifoldfree}) and the one read from the geometry
 (\ref{cubicconifoldcfocus}) manifest.

\section{Generalizations}
\label{secgen}

So far we focused on chiral quivers obtained by un-higgsing a non-chiral quiver with two gauge nodes, gauge group $SU(N)^2 \times U(1)_{diag}$ and the ABJM superpotential.
The quivers obtained in this way have $\sum k_i=0$ and we fixed CS levels to $\pm 1$ for the new gauge nodes generated by the un-higgsing. In this section we want to discuss generalizations of our results, by increasing the number of gauge nodes and the value of the CS levels.
Various generalizations are possible.
\begin{itemize}
\item
A first generalization of our analysis regards the cases of orbifolds with $k>1$. Following the rules of GK duality in such cases the main difference is that the original $U(N)_k$ nodes that undergo a duality become non abelian $U(|k|)_{-k}$ groups. At large $N$, keeping $k$ finite, the partition function is independent on these nodes as well. This is again, as explained in \cite{Jafferis:2011zi}, related to the fact that, treating $U(|k|)$ as a flavor symmetry, one can see that such  non-abelian flavor symmetry  is an accidental  flat direction of the free energy at large $N$. Then, a further gauging of the symmetry does not produce corrections at order $N^{3/2}$ in the free energy.

For example we can look at the quiver in Figure \ref{Fig:cubicele}, with CS levels
$k_3,k_4,k_5$ and $k_6$ such that\footnote{Other sign choices are possible and they do not spoil our conclusions.} $k_{3,4}>0$ and $k_{5,6}<0$ and 
$\sum_{i=3}^6 k_i = 0$.
The four dual gauge groups after applying \eqref{GKrule} become $\prod_{j=3}^{6} U(|k_i |)_{-k_i}$. At large $N$ once we evaluate the free energy for the rank $N$ gauge nodes we are effectively left with the partition function of pure $U(|k|)_{k}$
gauge theories which are fully gapped and contribute through a numerical factor to the free energy, that indeed does not affect the evaluation provided here.
 In addition the two nodes $1$ and $2$ in the dual quiver are $SU(N) \times SU(N) \times U(1)$, with vanishing CS level. 

We can compare this model with the one discussed in \cite{Jafferis:2011zi} by identifying \footnote{Where our choice sligltly differs from the one of \cite{Jafferis:2011zi}, that corresponds to  $k_3=n_{b_1}$, $k_4=n_{b_2}$, $k_5=-n_{a_1}$ and $k_6=-n_{a_2}$.} $k_3=n_{a_1}$, $k_4=n_{a_2}$, $k_5=-n_{b_1}$ and $k_6=-n_{b_2}$.
The constraint $n_{a_1}+n_{a_2} =n_{b_1}+n_{b_2} $ is satisfied automatically.
The dual superpotential is given in formula \eqref{cubicdual} and we can identify the fields $Y_{12}^{(i)}$  and $Y_{21}^{(i)}$ with the fields $a_i$ and  $b_i$ in \cite{Jafferis:2011zi} respectively.
The free energy for this model is 
\begin{equation}
F \!=\! \frac{2 \pi N^{3/2}}{3} \sqrt{
\prod_{j,k=1,2} \! \!(\Delta_{a_j} \!+\! \Delta_{b_k} ) 
\Big(\!\sum_{i=1}^2(n_{a_i} \Delta_{a_i} \!+\!n_{b_i} \Delta_{b_i})
\!-\!\frac{4\Delta_m^2}
 {(\sum_{i=1}^2(n_{a_i} \Delta_{a_i} \!+\!n_{b_i} \Delta_{b_i})}
\Big)
}
\end{equation}
We can extract the free energy for the electric theory by exploiting the duality dictionary implied by \eqref{GKrule} . Explicitly there are four relations associated to the bifundamentals.
They are
\begin{equation}
\begin{array}{ll}
\Delta_{Y_{12}^{(1)}} = \Delta_{X_{13}^{}} + \Delta_{X_{32}^{}},& \quad
\Delta_{Y_{12}^{(2)}} = \Delta_{X_{14}^{}} + \Delta_{X_{42}^{}},\\
\Delta_{Y_{21}^{(1)}} = \Delta_{X_{25}^{}} + \Delta_{X_{51}^{}},& \quad
\Delta_{Y_{21}^{(1)}} = \Delta_{X_{26}^{}} + \Delta_{X_{61}^{}}\, .
\end{array}
\end{equation}
In addition the effective FI of the dual theory $\tilde \Delta_m$ is related to the electric charges by 
$2 \tilde \Delta_m = 2 \Delta_m + k_3 (\Delta_{X_{13}} - \Delta_{X_{32}} )
+k_4 ( \Delta_{X_{14}} - \Delta_{X_{42}} )
-
k_5 (\Delta_{X_{25}} - \Delta_{X_{51}})
-
k_6 (\Delta_{X_{26}} - \Delta_{X_{61}})$.

We then obtain the free energy in the electric theory by using these relations.
We can also compare the result with the one expected from the geometry.
In this case the volume function can be written as
\begin{equation}
\label{expressionABJMfourflavors}
Vol (Y) \!=\! \frac{3\pi^4}{8}
 \frac{k_3 (\Delta _{\pi _1}+\Delta _{\pi _5})\!-\!k_5 (\Delta _{\pi _2}+\Delta _{\pi _6})+k_4 (\Delta _{\pi _3}+\Delta _{\pi _7})\!-\!k_6 (\Delta _{\pi _4}+\Delta _{\pi _8})}{
K
 (k_3 \Delta _{\pi _1}\! \! -\!k_5 \Delta _{\pi _2}\!+\!k_4 \Delta _{\pi _3}\!-\!k_6 \Delta _{\pi _4}) (k_3 \Delta _{\pi _5}\!-\!k_5 \Delta _{\pi _6}\!+\!k_4 \Delta _{\pi _7}\!-\!k_6 \Delta _{\pi _8})}
 \end{equation}
 where $\kappa_{i j k \ell} \equiv \Delta _{\pi _i}+\Delta _{\pi _j}+\Delta _{\pi _k}+\Delta _{\pi _\ell}$ and $K = \kappa _{1256} \kappa _{2367} \kappa _{1458} \kappa _{3478}$.
Indeed  we checked that this expression reproduces   the result obtained from the evaluation of the volumes in terms of the components of the Reeb vector.
 
We can parameterize the charge of each PM as 
\begin{equation}
\begin{array}{llll}
\Delta_{\pi_1} \!=\! \Delta_{X_{13}}\!-\!\frac{\Delta_{m}}{4 k_3}, &\quad
\Delta_{\pi_2} \!=\! \Delta_{X_{25}}\!+\!\frac{\Delta_{m}}{4 k_5}, &\quad
\Delta_{\pi_3} \!=\! \Delta_{X_{14}}\!-\!\frac{\Delta_{m}}{4 k_4}, &\quad
\Delta_{\pi_4} \!=\! \Delta_{X_{26}}\!+\!\frac{\Delta_{m}}{4 k_6},  \\
\Delta_{\pi_5} \!=\! \Delta_{X_{32}}\!+\!\frac{\Delta_{m}}{4 k_3}, &\quad
\Delta_{\pi_6} \!=\! \Delta_{X_{51}}\!-\!\frac{\Delta_{m}}{4 k_5},&\quad
\Delta_{\pi_7} \!=\! \Delta_{X_{42}}\!+\!\frac{\Delta_{m}}{4 k_4}, &\quad
\Delta_{\pi_8} \!=\! \Delta_{X_{61}} \!-\!\frac{\Delta_{m}}{4 k_6}\,.
\end{array}
\end{equation}
and observe that this parametrization makes the equivalence 
between the free energy read from the partition function and the one read from the geometry
 (\ref{expressionABJMfourflavors}) manifest.
Furthermore for $k_3=k_4=-k_5=-k_6=1$ formula (\ref{expressionABJMfourflavors})
 reduces to the one obtained for the cubic conifold, indeed in such case the numerator simplifies further because 
 $\sum_{i=1}^8 \Delta_{\pi_i} =2$.
 \item
In general any quiver with an arbitrary number of gauge nodes $G$, gauge group $SU(N)^G \times U(1)_{diag}$ and with a non-chiral like field content (not necessarily toric) gives rise to another quiver with a chiral like field content by un-higgsing some of the
bifundamentals. Decorating this last with CS levels that sum up to zero the corresponding free energy is expected to scale as $N^{3/2}$, by applying \eqref{GKrule} to the various un-higgsed nodes.
In the toric case this reflects into the fact that we expect a field theory that gives rise to the  $N^{3/2}$ scaling for any toric diagram without internal lattice point. Indeed only points on the faces or on the external edges are allowed if the toric diagram is obtained by lifting the PM of a non-chiral theory.
\item
In the discussion above we have restricted our interest to un-higgsed nodes with non-vanishing CS levels, where we have used the rules of the GK duality to show that the free energy reduces at large $N$ to the one of a quiver with non-chiral flavor introduced in \cite{Benini:2009qs}, that indeed scales as $N^{3/2}$ as proven in \cite{Jafferis:2011zi}.
We did not consider the case of vanishing CS level for the un-higgsed node, because the rule of the duality in this case is slightly different at the level of the three sphere partition function. Indeed, in this last case the dual model is obtained at physical level by Aharony duality \cite{Aharony:1997gp}, where a pair of gauge singlets, corresponding to electric monopoles acting as singlets in the dual phase should be introduced. Observe that this does not spoil our conclusion, because the dual gauge group vanishes in such case and the original non-chiral quiver is recovered.  Indeed the extra monopole do not contribute at large $N$ if the gauge group is $SU(N)^2 \times U(1)$. Again the chiral model obtained by un-higgsing and the original quiver have the same toric diagram and the toric duality persists also in this case at the level of the large $N$ free energy.
\item
If we consider $U(N)$ instead of $SU(N)$ gauge factors for a chiral quiver before the un-higgsing, the application of \eqref{GKrule}  introduces at finite $N$ also CS
levels  that are not properly quantized, which are not consistent with gauge invariance.
This does not spoil the large $N$ analysis, because at large $N$ their contribution is sub-leading but it is a problem to construct a proper dual phase.
The proper dual phase that we are looking for has indeed a flavor symmetry and not a $U(1)_{\pm 1}$ extra gauge node.
A consistent procedure to construct $U(N)$ chiral CS quivers dual $U(N)$ non-chiral quiver consist of applying the tensor deconfinement technique in presence of non vanishing CS levels. In such case the procedure acts directly on the CS levels and "deconfining" (see \cite{Berkooz:1995km}) a bifundamental gives rise to an $U(M)_k$ gauge node with $M \neq N$ even if of the same order.
The problem of this approach is that it cannot produces new cases with respect of the one already obtained at the level of the non-chiral quivers.
\end{itemize}

\subsection{Flavoring SPP:  a chiral model with five gauge nodes}

\begin{figure}[ht] \begin{center}
\includegraphics[width=6cm]{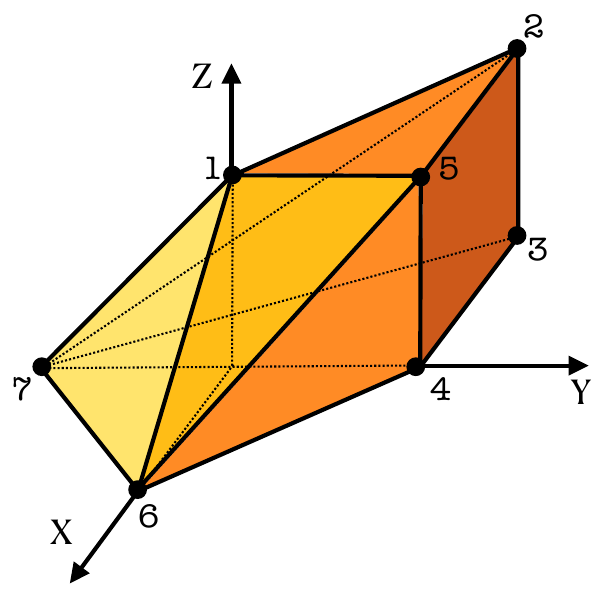}
\caption{Toric diagram for the flavored SPP. The same toric diagram has been 
obtained in \cite{Benini:2009qs} by flavoring the 4d quiver associated to the dP$_2$ singularity. One can indeed see the SPP toric diagram by a projection on the $(Z,Y)$ plane and the dP$_2$ model by a projection on the $(X,Y)$ plane. }
\label{spp1}
\end{center} \end{figure}

\begin{figure}[ht] \begin{center}
\includegraphics[width=8cm]{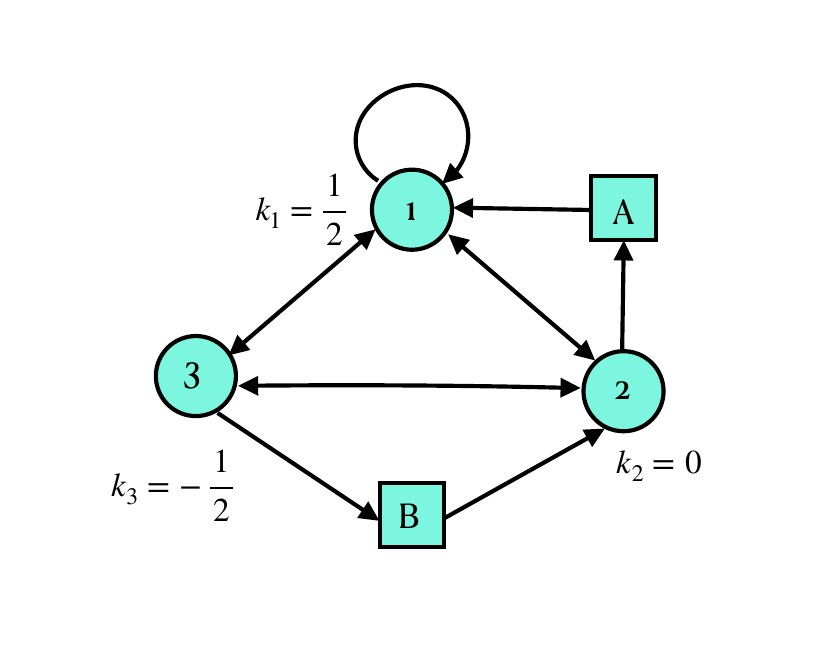}
\caption{Chiral flavoring of the SPP quiver associated to the superpotential  (\ref{SPPflavored}).}
\label{spp3}
\end{center} \end{figure}

As discussed above the un-higgsing prescription allows to construct chiral models associated to toric diagram without internal points. Some of these models can have also other description in terms of chiral quivers that do not have a known $N^{3/2}$ scaling. One example that we discuss here regards the toric diagram in Figure \ref{spp1}, which has appeared in \cite{Benini:2009qs} by a chiral flavoring of $dP_2$.
Actually this quiver gauge theory realization of this model does not satisfy the rules of \cite{Jafferis:2011zi}
for the cancellation of the long range forces and other quiver realization are needed.
\begin{figure}[ht] \begin{center}
\includegraphics[width=8cm]{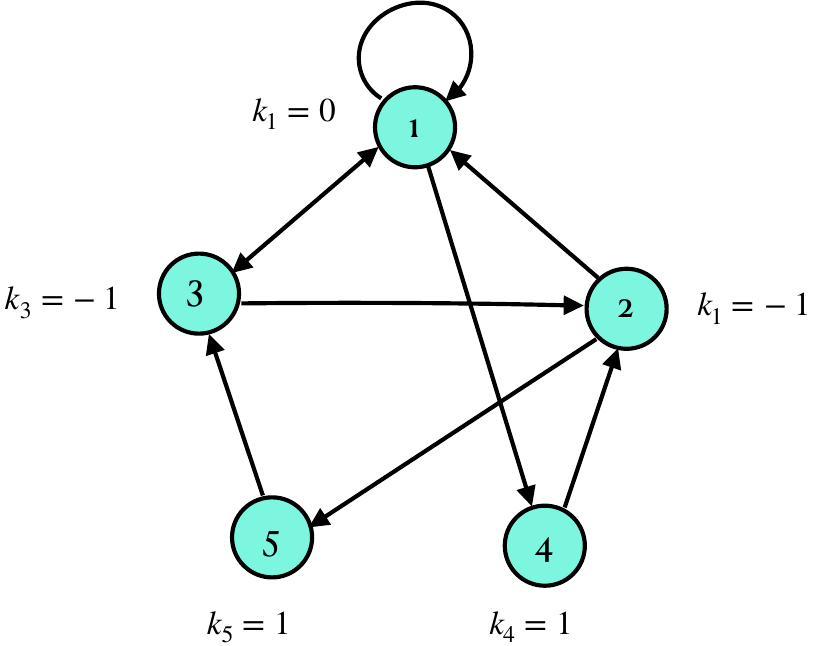}
\caption{Chiral quiver obtained by un-higgsing the SPP quiver. The associated superpotential is given by \eqref{Wun-higgsed}.}
\label{spp2}
\end{center} \end{figure}
The toric diagram in this case does not have internal points and it is then suitable to our analysis. In this case we can indeed project it on the plane $(y,z)$ and observe that on such  2d lattice we have the SPP toric diagram. Some of the points of the 3d toric diagrams are identified in the projection. There are three of such identifications, i.e. $v_3\leftrightarrow v_4$, $v_2\leftrightarrow v_5$ and 
$v_6 \leftrightarrow v_8$. While the last split is plausible for the SPP toric diagram by opportune CS levels assignations, because the point lies on the perimeter of the toric diagram,  the splits of the two external points is not possible \emph{per se}. It is then necessary to split the relative perfect matchings, i.e. to un-higgs the 4d quiver. The un-higgsed quiver and the CS assignation that reproduces the toric diagram are represented in Figure \ref{spp2}.
The superpotential for this model is given by
\begin{equation}
\label{Wun-higgsed}
W = X_{11} (X_{14}X_{42} X_{21} - X_{13} X_{31})  -  X_{14} X_{42} X_{25} X_{53} X_{32} X_{21} 
+ X_{25} X_{53}  X_{31} X_{13} X_{32} 
\end{equation}
The same quiver can be obtained by a chiral flavoring of the SPP model, where the quiver is shown in Figure \ref{spp3}.
In this case the model has superpotential
\begin{eqnarray}
\label{SPPflavored}
W &=& X_{11} (X_{12} X_{21} - X_{13} X_{31})  -  X_{12} X_{23} X_{32} X_{21} 
+ X_{23} X_{31} X_{13} X_{32} 
\nonumber \\
&+&
 X_{12} X_{2A} X_{A1} + X_{23} X_{3B} X_{B2} 
\end{eqnarray}
Again the validity of the large $N$ duality at the level of the partition function can be shown by considering the gauge group $SU(N_1) \times SU(N_2) \times SU(N_3) \times U(1)_{\text{diag}}$ while the un-higgsed nodes are $U(N_4)$ and $U(N_5)$.

We conclude the analysis of this model with an aside comment, that is non necessarily relevant for our analysis but that can be useful for future research direction.
The comment regards the study of the volume from the holographic dual perspective in terms of the charges associated to the toric divisors.

The volume is obtained as a function of the coordinates of the Reeb vector and it 
can be obtained from the $Z_{MSY}$ function as  $\text{Vol}(Y_7) = \frac{\pi^2}{12} Z_{MSY}$, where 
$Z_{MSY} = \sum_i \text{Vol}(\Sigma_i)$. The volumes $\text{Vol}(\Sigma_i)$ in this case are
\begin{eqnarray}
&&
\text{Vol}(\Sigma_1)= \frac{b_1-b_2+3 \left(b_3-4\right)}{\left(b_1+b_2-b_3+4\right) \left(b_3-4\right) \left(b_1+b_3-4\right) \left(b_1-b_2+b_3-4\right)} \nonumber \\ &&
\text{Vol}(\Sigma_2)=\frac{2 b_1+b_2-2 b_3+12}{\left(b_2-4\right) \left(2 b_1+b_2+4\right) \left(b_1+b_2-b_3+4\right) \left(b_3-4\right)}
 \nonumber \\ &&
\text{Vol}(\Sigma_3)=-\frac{2}{\left(b_2-4\right) \left(2 b_1+b_2+4\right) b_3}
\nonumber \\ &&
\text{Vol}(\Sigma_4)=\frac{1}{\left(b_2-4\right) \left(b_1+b_2-4\right) b_3}
 \nonumber \\ &&
\text{Vol}(\Sigma_5)= -\frac{b_1+b_2+b_3-8}{\left(b_2-4\right) \left(b_1+b_2-4\right) \left(b_3-4\right) \left(b_1+b_3-4\right)}
\nonumber \\ &&
\text{Vol}(\Sigma_6)=\frac{2 b_1+b_3-8}{\left(b_1+b_2-4\right) b_3 \left(b_1+b_3-4\right) \left(b_1-b_2+b_3-4\right)}
\nonumber \\ &&
\text{Vol}(\Sigma_7)=\frac{-3 b_1-3 b_2+b_3-12}{\left(2 b_1+b_2+4\right) \left(b_1+b_2-b_3+4\right) b_3 \left(b_1-b_2+b_3-4\right)}
\end{eqnarray}

The same volume can be obtained as a quartic function of the $\Delta_{\pi_i}$ charges  defined as
$\Delta_{\pi_i} = \frac{\pi}{6} \frac{\text{Vol}({\Sigma_i}}{Vol(Y_7)}$.
in terms of a quartic formula. 

In the examples found so far in the literature corrections to the generic term obtained by triangulating the toric diagram have been related to internal lines and internal faces. Here we observe that further corrections emerge in terms of internal volumes. Indeed the formula for the inverse $Z_{MSY}$  in this case is given by
\begin{eqnarray}
Z_{MSY}^{-1}&=&\frac{1}{6} \Big(
\sum_{i,j,k,l=1}^7 |\det(v_i,v_j v_k, v_\ell)| \Delta_i \Delta_j \Delta_k \Delta_\ell
-\frac{2}{3}  \big(\Delta _3^2 \Delta _1^2+\Delta _4^2 \Delta _1^2+\Delta _3 \Delta _4 \Delta _1^2\nonumber\\
& -&2 \Delta _2 \Delta _4 \Delta _6 \Delta _1-\Delta _2 \Delta _3 \Delta _6 \Delta _1-\Delta _3 \Delta _5 \Delta _7 \Delta _1-2 \Delta _4 \Delta _5 \Delta _7 \Delta _1
\nonumber\\
&+ &\Delta _2^2 \Delta _6^2+\Delta _5^2 \Delta _7^2-\Delta _2 \Delta _5 \Delta _6 \Delta _7\big) \Big)
\end{eqnarray}
Observe that the corrections in this case are all expressed by combining the internal lines, i.e. the lines connecting $v_1$ with either $v_3$ or $v_4$, the line connecting $v_2$ with $v_6$ and the line connecting $v_5$ and $v_7$.
By inspection we have observed the same type of corrections by exploring other various 3D toric  diagrams, i.e. the corrections come by combining in all the possible ways the $\Delta_i$ that correspond to  endpoints of lines that do not lie on any face of the toric diagram. Some corrections to the quartic formula correspond then to the ones already observed in \cite{Amariti:2011uw,Amariti:2012tj} but more general possibilities are allowed by more sophisticated geometries, as the one studied here.

\begin{figure}[H] 
\begin{center}
\includegraphics[width=7cm]{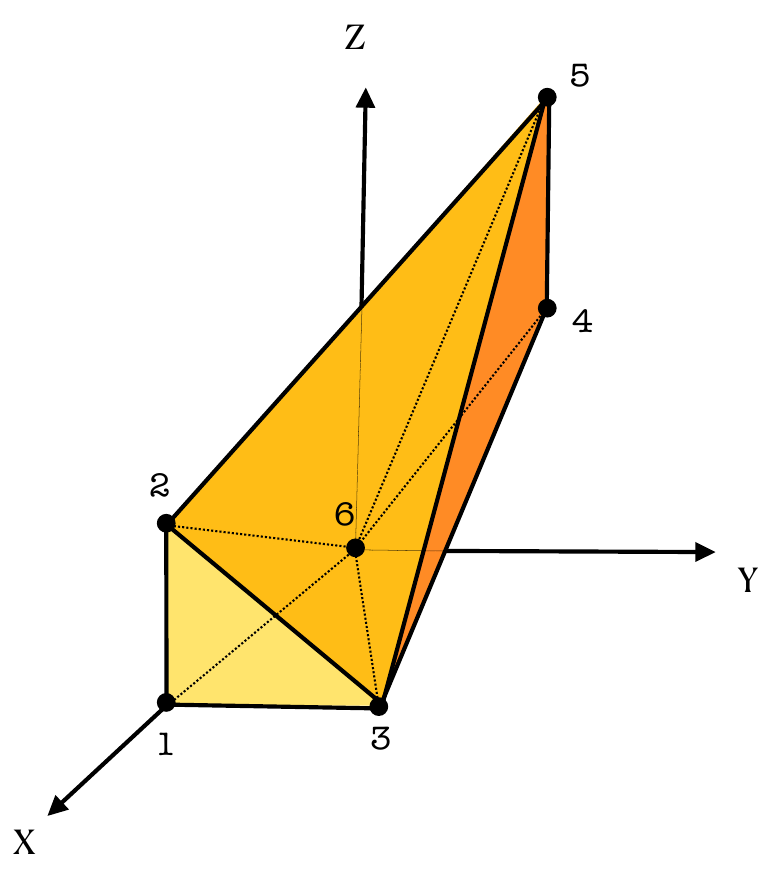}
\caption{Toric diagram for the Q$^k$ family.}
\label{Qkfig}
\end{center} \end{figure}

As an  additional example of this type we consider  the toric diagram of the Q$^k$ family, This model either correspond to the chiral flavored quiver studied in \cite{Jafferis:2011zi} or it can be studied along the lines of the discussion above, by considering the same un-higgsed quiver studied in sub-section \ref{subsec:CtimesC} with the CS levels assigned as in Figure Figure \ref{CtimesConifold}, for generic $k>1$.
Here we skip the detailed analysis of the models and directly focus on the structure of the quartic formula.
the toric diagram in this case is
\begin{equation}
G_t^{(Q^k)} = \left(
\begin{array}{cccccc}
 1 & 1 & 1 & 0 & 0 & 0 \\
 0 & 0 & 1 & 1 & 1 & 0 \\
 0 & 1 & 0 & \kappa -1 & \kappa  & 0 \\
 1 & 1 & 1 & 1 & 1 & 1 \\
\end{array}
\right)
\end{equation}
and by inspection we found that the quartic formula that reproduces the inverse $Z_{MSY}$ function 
is
\begin{eqnarray}
Z_{MSY}^{-1} &=&  
\frac{1}{6} \sum_{i,j,k,l=1}^6 |\det(v_i,v_j v_k, v_\ell)| \Delta_i \Delta_j \Delta_k \Delta_\ell
-
\frac{4 }{\kappa+1}
(2 \Delta _4^2 \Delta _1^2 (\kappa -1)+\Delta _5^2 \Delta _1^2 \kappa 
\nonumber \\
&+&
2 \Delta _4 \Delta _5 \Delta _1^2 (\kappa -1)+2 \Delta _2 \Delta _4^2 \Delta _1 (\kappa -1)+\Delta _2^2 \Delta _4^2 \kappa -2 \Delta _2 \Delta _4 \Delta _5 \Delta _1)
\end{eqnarray}
Again we observe that there are three internal lines in the geometry, connecting  the points $v_1-v_4$, $v_2-v_ 5$ and $v_2-v_4$ respectively.
Again all the possible combinations of these lines appear in the quartic formula, in analogy with the case of the flavored/un-higgsed SPP discussed above.
It should be interesting to predict the correction coefficients by finding their physical origin. We leave such problem to future investigations.

\section{Discussion}
\label{discussion}

In this paper we study the large $N$ scaling of the free energy for 3d chiral quivers obtained by un-higgsing 4d grand-parents, that correspond to quivers describing  D3 branes probing the tip of toric CY$_3$ manifolds with a SE$_5$ base. The explicit examples studied  in Section \ref{csecex} consist of un-higgsing the 4d conifold in various ways, and in addition we consider on the chiral quiver a CS assignation compatible with the holographic correspondence. In this way the moduli space of the 3d quivers are indeed associated to M2 branes probing the tip of toric CY$_4$ manifolds with a SE$_7$ base. 
The moduli space of the models considered here can be equivalently obtained from a different approach, consisting of a chiral flavoring of the 4d grand-parent quiver, with possible addition of CS terms.
The relation between these two classes of models is stronger that just a correspondence among their moduli space; indeed, by exploiting exact relations among hypergeometric hyperbolic integrals one can see that the models are IR dual, where the duality generalizes the one obtained originally in \cite{Giveon:2008zn} (see also \cite{Aharony:2008gk}). For this reason we referred here to such duality as Giveon-Kutasov duality.
Once one focuses on the large $N$ evaluation of the partition function the duality predicts the $N^{3/2}$ scaling for the chiral quivers, confirming the results recently obtained numerically in \cite{Hosseini:2025jxb} and extending the expectation to various other models.
Even if here we restricted to the un-higgsing of the conifold, our result is more general and it holds for any other un-higgsing of a 4d non-chiral quiver (non-necessarily toric). Reversing the argument, using the  inverse algorithm presented in the paper,  we claim that the chiral quivers whose $N^{3/2}$ scaling can be proven with our analysis have a toric diagram without internal points. 
While many toric diagrams can be captured from this construction there are many others that are beyond this class. Two famous examples are M$^{111}$ and $Q^{222}$. Interestingly, as discussed in \cite{Hosseini:2025jxb}, in this case there is no numerical evidence of the $N^{3/2}$ scaling from the chiral quiver description as well.
\\

Various extensions of our work are possible.
Here we provided examples associated to the ones studied explicitly in \cite{Jafferis:2011zi}, that correspond to the flavoring of the conifold quiver. We leave a generalization of the analysis for flavored quivers with more than two gauge groups to future studies.
Observe also that, while the leading contribution to the three sphere free energy for the models considered in this paper is equivalent to the one expected for the flavored quivers, sub-leading corrections are in general different, due to the extra $U(|k_i|)_{k_i}$ gauge nodes. It would be interesting to have a systematic way to compute such corrections. An useful work in this direction is  \cite{Geukens:2024zmt}.
We also proposed a generalization of the structure of the quartic formula that reproduces the volume computation.  The quartic formula is also useful in the calculation of the master volume \cite{Gauntlett:2018dpc}, because it allows to construct  an entropy function for AdS$_4$ BPS black holes in M-theory with general magnetic charges. However, as recently proven in  \cite{Hosseini:2025ggq,Hosseini:2025mgf}, when considering baryonic twists, the quartic expression has to be modified properly. It should be interesting to provide the explicit expression for flavored/un-higgsed SPP and the Q$^k$ family discussed here.

\section*{Acknowledgments}
We are grateful to Alberto Zaffaroni for discussions.
This work  has been supported in part by the Italian Ministero dell'Istruzione, Università e Ricerca (MIUR), in part by Istituto Nazionale di Fisica Nucleare (INFN) through the “Gauge Theories, Strings, Supergravity” (GSS) research project.

\appendix

\section{$\mathcal{N}=2$ CS toric quivers and the moduli space}
\label{toricdiagram}
In this appendix we briefly review the main aspects of 3d toric quiver gauge theories
and to the study of their moduli space.
These models are three dimensional CS quiver gauge theories, which are believed to represent 
the low energy dynamics of a stack of M2 branes at a toric CY$_4$ singularity.
The gauge fields do not have any kinetic term  but they have CS terms.
These theories are conjectured to be $\mathcal{N}=2$ supersymmetric field theories with a product
gauge group $\prod_{a=1}^G U(N_a)$, where $G$ is the total amount of gauge groups, and matter field in the bifundamental or in the adjoint representation of the gauge groups.
These theories are described by the $\mathcal{N}=2$ Lagrangian
\begin{eqnarray}
 \label{Lag}
\mathcal{L} &=& 
-i\sum_a k_a \int_0^1 dt V_a \overline D^\alpha (e^{t V_a} D_\alpha e^{-t V_a})
\nonumber \\
&-&
\int d^4 \theta \sum_{X_{ab}} X^{\dagger}_{ab} e^{-V_a} X_{ab} e^{t V_a}
+\int d^2 \theta W(X_{ab})+c.c.
\end{eqnarray}
where the $V_a$ are the vector superfields associated to the $U(N_a)$ gauge group and the fields $X_{ab}$ are the matter chiral superfields in the bifundamental or in the adjoint representation
of the gauge groups.
$D_\alpha$ is the derivative operator in the superspace, the $k_a $ are the CS levels and $W$ is the superpotential.
As we anticipated above we study toric Calabi-Yau fourfolds. The toric condition is translated in the bipartite structure of the superpotential. Indeed every bifundamental
field  appears twice in the superpotential, one time with the plus and one other with the minus sign in the coupling constant.
The vector field $V_a$ has a real scalar component that originates from the dimensional reduction of the fourth component of the four dimensional
$A_\mu$ field.
In component notation the 3d vector multiplet  is  $V_a=(A_\alpha,\chi_{\alpha}, \sigma_\alpha,D_\alpha)$, and the CS term in (\ref{Lag}), in the WZ
gauge,  is explicitly written as
\begin{equation}
S_{CS} = \sum_a \frac{k_a}{4 \pi} \int \Tr \left(
A_\alpha \wedge \text{d} A_{\alpha}+\frac{2}{3} A_\alpha \wedge A_\alpha \wedge A_{\alpha} 
-\overline \chi_\alpha \chi_\alpha + 2 D_\alpha \sigma_\alpha 
\right)
\end{equation}
The classical moduli space of these theory is studied as in four dimensions. The moduli space indeed is given by the vanishing condition on the scalar potential.
They reduce to the solution of the equation of motion, namely the $F$ and the $D$ terms.
These equations are
\begin{eqnarray} 
\partial_{X_{ab}} W &=&0 \label{EOM1}\\
\mu_a(X)=\sum_b X_{ab}X_{ab}^{\dagger} - \sum_c X_{ca}^{\dagger} X_{ca}
+[X_{aa},X_{aa}^{\dagger} ]&=&4 k_a \sigma_a \label{EOM2}\\
\sigma_a X_{ab} - X_{ab} \sigma_b&=&0 \label{EOM3}
\end{eqnarray} 
The first equation is the solution of the $F$ terms condition coming from the superpotential. The second equation is analog to the four dimensional D-term. However the three dimensional
 D-term are more involved than in four dimensions. Indeed in this case there is a new equation (\ref{EOM3}), coming from the
 scalar component of the vector multiplet.

In this paper we study the moduli space for $N_a=1$. This is the  mesonic moduli space $\mathcal{M}$ corresponding to the 
space transverse to a single M2-brane.  It  is a four dimensional toric Calabi-Yau
cone. For higher $N_a$  $\mathcal{M}$ is the N-th symmetric product of the one
for $N_a = 1$, which is not toric anymore \cite{Forcella:2008bb}. For this reason we just study the toric data for the case $N=1$.

By solving the equation of motion it was observed \cite{Martelli:2008si,Hanany:2008cd} that there is a constraint on the
CS levels
\begin{equation}
\sum_{a=1}^G N_a k_a = 0 
\end{equation}
otherwise the mesonic moduli space is three and not four dimensional. 
Moreover from equation (\ref{EOM3}) we have $\sigma_a=\sigma$. 
Equation (\ref{EOM1}) gives the space of solution of the $F$ terms, i.e. the master space,
which in this case is a toric variety of dimension $G+2$, where $G$ is the number of gauge groups.
Then the moduli space is obtained by modding this space by the D-term constraints, which in this case are given by the remaining $U(1)^{G-2}$. Indeed, differently from the four dimensional case where the modding is done by $G-1$ abelian factors, here there is an extra condition coming from the CS levels. This extra condition is  fixed by the choice of $\sigma$. By defining
$~^{Irr}\mathcal{F}^{\mathbf{b}}$ the coherent component of the master space, the
mesonic moduli space is formally given by
\begin{equation}
\mathcal{M} = ~^{Irr}\mathcal{F}^{\mathbf{b}}/H
\end{equation}
where $H$ is the $(\mathbb{C}^*)^{G-2}$ kernel of 
\begin{equation}
C = \left(\begin{array}{cccccc}
1&1&1&1&1&1\\
k_1&k_2&\dots&\dots&k_{G-1}&k_G
\end{array}
\right)
\end{equation}
In the next section we will show an algorithm to compute this moduli space from toric geometry.
Indeed, as in four dimensions,  it is possible to encode the solution of the classical equation of motion in a geometrical lattice, called toric diagram.

\subsection{Computation of the toric diagram}

There are many algorithms to compute the toric data for a given CS toric quiver gauge theory.
In this paper we used an algorithm given by a mixture of the algorithms of \cite{Hanany:2008gx} and
of \cite{Ueda:2008hx}.
One first computes the diagram for a 4D theory by looking at the solution of the D-terms and of the F-terms and then adds the contribution due to the CS terms.
The F-term solutions are given by the perfect matching matrix. 
The perfect matchings of a toric gauge theory are sets of fields each one appearing
just once in every superpotential term. They are associated with the points of the toric diagram.
The D-terms are taken into account by the charged of the GLSM associated to the 
quiver gauge theory. By defining $d$ the adjacency matrix and $P$ the perfect matching matrix, 
the GLSM matrix is defined as $Q$, given by the relation $d= Q \cdot P^T$.
Then the D-terms are given by modding the $Q$ matrix by the gauge transformation. 
The four dimensional moduli space is encoded in toric diagram given by
\begin{equation}
\label{TD}
G_T^{(4D)} = \left(
\begin{array}{c}
Q_F\\
Q_D
\end{array}
\right)=
\left(
\begin{array}{c}
Ker(P)\\
Ker(1_1,\dots,1_G) Q
\end{array}
\right)
\end{equation}
The toric diagram for the parent CS toric quiver in three dimensions is  computed from (\ref{TD}) by adding the CS levels contribution to the $P$ matrix.
This is done by assigning a flux of CS levels to each field in the quiver, such that the total flux is conserved. These fluxes are then taken into account as an additional row in the $P$ matrix, where each entry equals the sum over the flux contributions of the fields within the corresponding column.

\subsection{Example: ABJM}
\label{primoex}
The superpotential is
\begin{equation}
\label{spotconi}
W = X_{12} X_{21} Y_{12} Y_{21} - X_{12} Y_{21} Y_{12} X_{21}
\end{equation}
where the  gauge group is  $U(N)_k \times U(N)_{-k}$. 
In the quiver we associated a 
conserved flux of CS levels to every field.
The adjacency matrix and the perfect matching matrix are 
\begin{eqnarray}
 \label{PMconi}
d_{ABJM} =
\left(
\begin{array}{c||cccc}
&X_{12}&Y_{12}&X_{21}&Y_{21}\\
\hline
\hline
N_1&1&1&-1&-1\\
N_2&-1&-1&1&1
\end{array}
\right) 
\quad, \quad
P_{ABJM}  =
\left(
\begin{array}{c||cccc}
X_{12}&1&0&0&0\\
Y_{12}&0&1&0&0\\
X_{21}&0&0&1&0\\
Y_{21}&0&0&0&1\\
\hline
\hline
\text{CS}&0&0&0&k
\end{array}
\right) 
\end{eqnarray}
The matrix $Q$ is the same as $d$ because $P= \mathbf{1}_4$.
The 4D toric diagram is
\begin{equation}
 \label{torABJM}
G_t = \left(
\begin{array}{c}
Q_F\\
Q_D
\end{array}
\right)
= Ker\left(
\begin{array}{c}
Ker(P)\\
Q
\end{array}
\right)= Ker[Q]=
\left(
\begin{array}{cccc}
 1 & 1 & 1 & 1 \\
 1 & 0 & 1 & 0 \\
 0 & 1 & 1 & 0
\end{array}
\right)
\end{equation}
The 3D diagram is found by adding the CS row
to (\ref{torABJM}).

\section{Non-chiral flavor and un-higgsing of adjoints}
\label{onegroup}

In this appendix we discuss the generalization of our analysis to cases with non-chiral flavors. The models under investigation have been studied in \cite{Benini:2011mf} and they have a single $U(N)$ gauge group, three adjoints $\Phi_{1,2,3}$
and three $SU(f_{i=1,2,3})$  flavor groups with superpotential
\begin{equation}
\label{B1}
W = \phi_1 [\phi_2, \phi_3] + \sum_{i=1}^{3} q_i \phi_i \tilde q_i
\end{equation}
where the fundamentals $q_i$ and the antifundamentals $\tilde q_i$ are rotated under the $SU(f_i)$ flavor symmetry.

In this case the CS level is vanishing and the quantum constraint on the monopoles is $T \tilde T = \prod_{i=1}^{3} \phi_i^{f_i}$.
The toric diagram is build from the six lattice points
\begin{eqnarray}
\label{genfree}
&&
v_{1} = (0,0,0),\quad
v_{2} = (0,1,0),\quad
v_{3} = (1,0,0),\nonumber \\
&&
v_{4} = (0,0,f_1),\quad
v_{5} = (0,1,f_2),\quad
v_{6} = (1,0,f_3) \, .
\end{eqnarray}
The free energy has been computed in \cite{Jafferis:2011zi} and it is 
\begin{equation}
\label{freen1n2n3}
F_{S^3} =\frac{2\sqrt2 \pi N^{3/2}}{3}
\sqrt{\Delta_{1} \Delta_{2} \Delta_{3} 
\left(
\sum_{i=1}^3 f_i \Delta_i -\frac{4\Delta_m^2}{\sum_{i=1}^3 f_i \Delta_i }
\right)
}
\end{equation}
In the following we show that  the non-chiral quivers obtained by un-higgsing either one or two or three  adjoints 
have the same free energy obtained from the flavored models just reviewed.
    \begin{figure}[H] \begin{center}
\includegraphics[width=10cm]{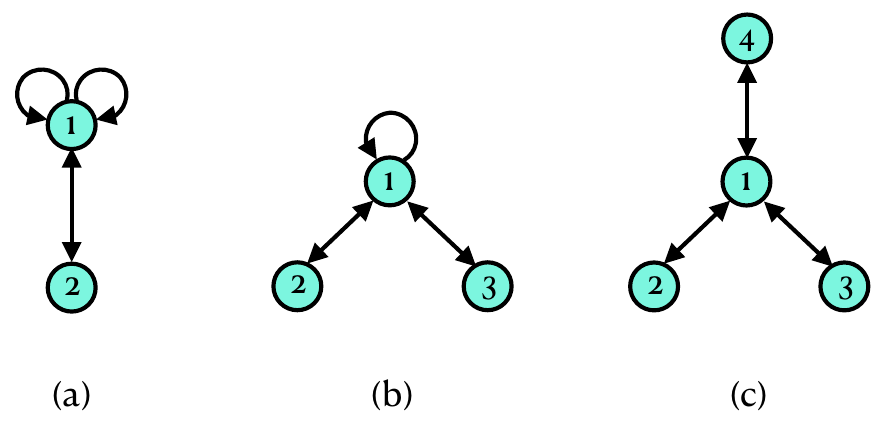}
\caption{The three possible un-higgsings of the 1-node quiver gauge theory with three adjoint fields.}
\label{n1n2n3quiver}
\end{center} \end{figure}

\begin{itemize}
\item We start by un-higgsing a single adjoint. This corresponds to the dual ABJM model originally studied in \cite{Hanany:2008fj}. In this case we have an $U(N)_1 \times U(N)_{-1}$ gauge theory with two adjoints $\phi_{1,2}$ and two bifundamentals $a,b$. 
The quiver is shown in Figure \ref{n1n2n3quiver} (a) and it has superpotential
\begin{equation}
W  = a b [\phi_1,\phi_2]
\end{equation}
The free energy of this model has been studied from the operator counting formalism \cite{Gulotta:2011si,Gulotta:2011aa,Gulotta:2011vp} in \cite{Kim:2012vza}.

Here instead we apply formula \eqref{GKrule} on the three sphere free energy to obtain the model with $f_1=f_2=0$ and $f_3=1$ discussed above, with in addition an extra $U(1)_{-1}$ gauged flavor symmetry.
The FI $\tilde \Delta_{m_1}$ of the $U(N)_0$ gauge group is $\tilde \Delta_{m_1}  = \Delta_{m_1} +\Delta_{m_2} +  (\Delta_a-\Delta_b)/2$ and 
the adjoint $\Phi_3$ has $R$-charge $\Delta_a + \Delta_b$.
The leading $N^{3/2}$ contribution to the free energy for this dual phase is obtained from (\ref{freen1n2n3}) and it is explicitly
\begin{equation}
\label{freendualABJM}
F_{S^3} =\frac{2\sqrt2 \pi N^{3/2}}{3}
\sqrt{\Delta_{1} \Delta_{2} \Delta_{3} 
\left(
 \Delta_3 -\frac{4\tilde \Delta_{m_1}^2}{\Delta_3 }
\right)
}
\end{equation}
It follows that the free energy for the quiver with two rank N gauge nodes can be expressed in terms of the charges of the fields $a,b,\phi_1$ and $\phi_2$ as 
\begin{equation}
\label{freendualABJM2}
F_{S^3} =\frac{4\sqrt2 \pi N^{3/2}}{3}
\sqrt{\Delta _1 \Delta _2 \left(\Delta _a+\Delta _m\right) \left(\Delta _b-\Delta _m\right)}
\end{equation}
where $\Delta_m = \Delta_{m_1} +\Delta_{m_2}$.
This is consistent with the results of \cite{Kim:2012vza} upon setting $\Delta_a = \Delta_b$.
\item The second example correspond to the toric diagram with $f_2=f_3=1$ and $f_1=0$ and it is obtained by un-higgsing two adjoints.
The model is an $U(N)_1 \times U(N)_1 \times U(N)_{-2}$ quiver and it was originally studied in \cite{Davey:2009sr} (see also \cite{Benishti:2009ky}).
The quiver is shown in Figure \ref{n1n2n3quiver} (b) and it has superpotential
\begin{equation}
\label{spotcc}
W  =\phi_1 [ab,cd]
\end{equation}
If we apply formula \eqref{GKrule} on the three sphere free energy, we obtain the model with $f_2=f_3=1$ and $f_1=0$ discussed above, with in addition an extra $U(1)_{-1}^2$ gauged flavor symmetry.
The FI is of the $U(N)_0$ gauge group is  $\tilde \Delta_{m_1}  =\Delta_{m_1}  +\Delta_{m_2}  +\Delta_{m_3}+( \Delta_a-\Delta_b+\Delta_c-\Delta_d)/2$ and 
the adjoints $\Phi_2$  and  $\Phi_3$ have $R$-charges $\Delta_a + \Delta_b$ and $\Delta_c + \Delta_d$ respectively.
The leading $N^{3/2}$ contribution to the free energy for this dual phase is obtained from (\ref{freen1n2n3}) and it is explicitly
\begin{equation}
\label{dueflavors}
F_{S^3} =\frac{2\sqrt2 \pi N^{3/2}}{3}
\sqrt{\Delta_{1} \Delta_{2} \Delta_{3} 
\left(
 \Delta_2+ \Delta_3 -\frac{4\tilde \Delta_{m_1}^2}{\Delta_2+\Delta_3 }
\right)
}
\end{equation}
The free energy for the quiver with three rank N gauge nodes can be expressed in terms of the charges of the fields $a,b,c,d$ and $\phi_1$ as 
\begin{equation}
F_{S^3} =\frac{4\sqrt 2 \pi N^{3/2}}{3}
 \sqrt{\frac{\Delta _1 \left(\Delta _a+\Delta _b\right) \left(\Delta _c+\Delta _d\right) \left(\Delta _a+\Delta _c+\Delta _m\right) \left(\Delta _b+\Delta _d-\Delta _m\right)}{\Delta _a+\Delta _b+\Delta _c+\Delta _d}}
\end{equation}
where $\Delta_m= \sum_{i=1}^3 \Delta_{m_i}$.
This result  corresponds to the expectation from the volume computation of $\mathbb{C} \times \mathcal{C} $
given in formula (\ref{T11Cvol}) once the superpotential constraint following from  \eqref{spotcc} is considered
\item 
The  third possibility consists of having $f_{1,2,3}=1$. This case can be reproduced by a four groups non-chiral quiver, obtained by un-higgsing the three adjoints.
The model is an $U(N)_1^3 \times U(N)_{-3}$ quiver and it was originally studied in \cite{Hanany:2008fj}.
The quiver is shown in Figure \ref{n1n2n3quiver} (c) and it has superpotential
\begin{equation}
\label{spot3D3}
W  =ab [cd,ef]
\end{equation}
If we apply formula \eqref{GKrule} on the three sphere free energy we obtain the model with $f_{1,2,3}$  discussed above, with in addition an extra $U(1)_{-1}^3$ gauged flavor symmetry.
The FI is of the $U(N)_0$ gauge group is $\tilde \Delta_{m_1} =\sum_{a=1}^4 \Delta_{m_a} +( \Delta_a-\Delta_b+\Delta_c-\Delta_d+\Delta_e-\Delta_f)/2$ and 
the adjoints $\Phi_{1,2,3}$ have $R$-charges $\Delta_a + \Delta_b$, $\Delta_c + \Delta_d$ and $\Delta_e + \Delta_f$  respectively.
The leading $N^{3/2}$ contribution to the free energy for this dual phase is obtained from (\ref{freen1n2n3}) and it is explicitly
\begin{equation}
\label{treflavors}
F_{S^3} =\frac{2\sqrt2 \pi N^{3/2}}{3}
\sqrt{\Delta_{1} \Delta_{2} \Delta_{3} 
\left(
 \Delta_1+\Delta_2+ \Delta_3 -\frac{4\tilde \Delta_{m_1}^2}{ \Delta_1+\Delta_2+\Delta_3 }
\right)
}
\end{equation}
The free energy for the quiver with four rank N gauge nodes can be expressed in terms of the charges of the fields $a,b,c,d,e$ and $f$ as 
\begin{eqnarray}
\label{finaletregaugings}
F_{S^3} =\frac{4 \pi N^{3/2}}{3}
&& \sqrt{\Delta _a+\Delta _b) (\Delta _c+\Delta _d) (\Delta _e+\Delta _f)}\nonumber 
 \\ \times &&
 \sqrt{ (\Delta _a+\Delta _c+\Delta _e+\Delta _m) (\Delta _b+\Delta _d+\Delta _f-\Delta _m)}
\end{eqnarray}
where $\Delta_ m =\sum_{a=1}^4 \Delta_{m_a}$ and where we have 
imposed the constraints imposed by the superpotential \eqref{spot3D3}
on the denominator of (\ref{finaletregaugings}).  We conclude observing that (\ref{finaletregaugings})
coincides with the one result expected from the volume computation in formula (\ref{D3vol}).
\end{itemize} 
We can generalize the analysis to higher CS levels, by considering a four nodes quiver with gauge group   $U(N)_{-(k_1+k_2+k_3)} \times \prod_{i=1}^{3} \times U(N)_{k_i}$ gauge theory, 
where the first gauge group is connected by a pair of conjugate bifundamentals to each other gauge group and the superpotential is \eqref{spot3D3}.
By \eqref{GKrule} the  free energy of this theory is equivalent to the one of an $U(N)_0 \times \prod_{i=1}^{3} \times U(|k_i|)_{-k_i}$ gauge theory, with three pairs of conjugated bifundamentals as above and in addition three $U(N)$ adjoints. The superpotential corresponds to the one in \eqref{B1}, where $\phi_i$ are the adjoints corresponding to the mesons formed by the conjugated bifundamentals for each gauge node that undergoes a duality and $q_i$ and $\tilde q_i$ are the dual 
 conjugated bifundamentals.
 The duality dictionary  in this case gives 
\begin{eqnarray}
\label{doic}
&&
\Delta_{1} = \Delta_{a} + \Delta_{b}\,, \quad \quad
\Delta_{2} = \Delta_{c} + \Delta_{d}\,, \quad \quad
\Delta_{3} = \Delta_{e} + \Delta_{f}\, ,\nonumber \\
&&
2\tilde \Delta_{m_1} = 2\Delta_{m} +|k_1|(\Delta_{a} - \Delta_{b})+|k_2|( \Delta_{c} - \Delta_{d})+|k_3|( \Delta_{e} - \Delta_{f})\, .
\end{eqnarray}
  Then. plugging (\ref{doic}) in \eqref{freen1n2n3} we obtain 
  \begin{equation}
  \label{dualfree}
F_{S^3} =\frac{4\sqrt2 \pi N^{3/2}}{3}\sqrt{
    \frac{
  K
    (|k_1| \Delta _a\!+\!|k_2| \Delta _c\!+\!|k_3| \Delta _e \!+\! \Delta _m ) 
    (|k_1| \Delta _b\!+\!|k_2| \Delta _d\!+\!|k_3| \Delta _f  \!-\! \Delta _m)
}{ |k_1| (\Delta _a\!+\!\Delta _b)\!+\!|k_2| (\Delta _c\!+\!\Delta _d)\!+\!|k_3| (\Delta _e\!+\!\Delta _f)}}
  \end{equation}
  where $K=   (\Delta _a\!+\!\Delta _b) 
    (\Delta _c\!+\!\Delta _d) 
    (\Delta _e\!+\!\Delta _f) $ and $\Delta_m = \sum_{a=1}^4 \Delta_{m_a}$.
    
    \begin{figure}[ht] \begin{center}
\includegraphics[width=6cm]{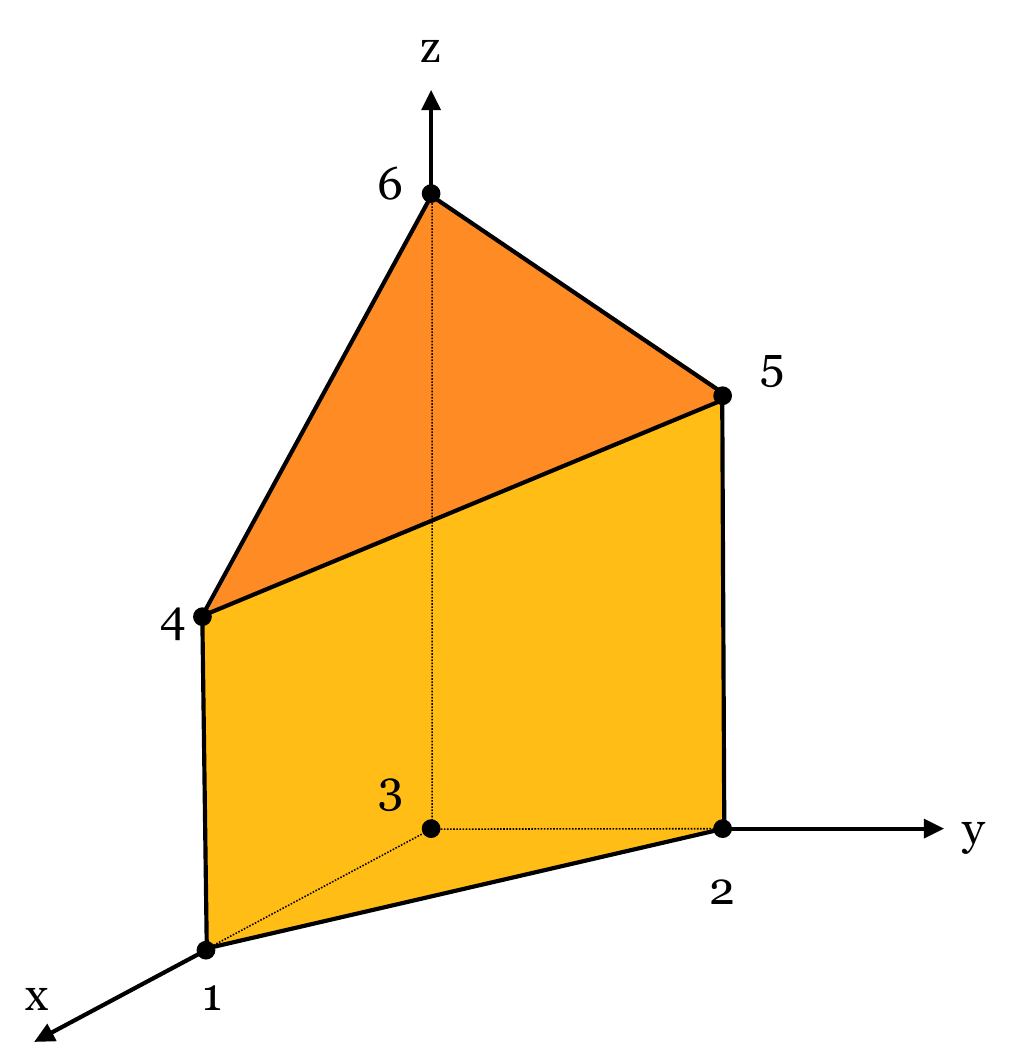}
\caption{Toric diagram for the flavored $U(N)$ model with three adjoints, or equivalently the non-chiral quiver with four nodes.}
\label{n1n2n3}
\end{center} \end{figure}
    We conclude by comparing the free energy just obtained for the non-chiral quiver using \eqref{GKrule}
    with the free energy obtained from the volume computation in the dual setup.
    In this case the toric geometry can be represented in Figure \ref{n1n2n3} where the lattice points and the fields in the associated
    PM are 
\begin{equation}
\label{toricD3app}
G_{t}^{D_3}
=
\left(
\begin{array}{cccccc}
 1 & 0 & 0 & 1 & 0 & 0 \\
 0 & 1 & 0 & 0 & 1 & 0 \\
 0 & 0 & 0 & |k_1| & |k_2| & |k_3| \\
  1 & 1 & 1 & 1 & 1 & 1 \\
  \hline
a & c & e& b &d&f \\
\end{array}
\right)
\end{equation}
The $Z_{MSY}$ function for this geometry can be shown to be  equivalent to the expression
  \begin{equation}
  \label{volfree}
  Vol(Y) = \frac{\pi^4}{12}
    \frac{  \sum_{i=1}^3 |k_i| \Delta _{\pi_i} +\sum_{i=1}^3 |k_i| \Delta _{\pi_{i+3}}  }
    { (\Delta _{\pi_1}+\Delta _{\pi_4}) 
    (\Delta _{\pi_2}+\Delta _{\pi_5}) 
    (\Delta _{\pi_3}+\Delta _{\pi_6}) 
    (\sum_{i=1}^3 |k_i| \Delta _{\pi_i})  (\sum_{i=1}^3 |k_i| \Delta _{\pi_{i+3}})  
}
  \end{equation}
once the charges $\Delta_{\pi_i} = \frac{2 \text{Vol}(\Sigma_i)}{Z_{MSY}}$ are 
 computed in terms of the components of the Reeb vector.   
  The charges of the PM can be expressed in terms of the charges of the fields in the non-chiral quiver as
  \begin{eqnarray}
  \label{parpmfields}
&&
\Delta_{\pi_1} = \Delta_a+\frac{\Delta_m}{3 |k_1|}\, , \quad \quad 
\Delta_{\pi_2} = \Delta_c+\frac{\Delta_m}{3 |k_2|}\, , \quad \quad
\Delta_{\pi_3} = \Delta_e+\frac{\Delta_m}{3 |k_3|}\, , 
\nonumber \\
&&
\Delta_{\pi_4} = \Delta_b-\frac{\Delta_m}{3 |k_1|}\, , \quad \quad
\Delta_{\pi_5} = \Delta_d-\frac{\Delta_m}{3 |k_2|}\, , \quad \quad
\Delta_{\pi_6} = \Delta_f-\frac{\Delta_m}{3 |k_3|}\, .
\end{eqnarray}
Using this parameterization, the equivalence between the free energy  \eqref{dualfree} and the geometric free energy obtained from the volume \eqref{volfree} is manifest.
    
\bibliographystyle{JHEP}
\bibliography{ref32.bib}
\end{document}